%% file: HeisenbergMomentosUR.tex
\documentclass[aps,pra,preprint,superscriptaddress,showpacs,showkeys]{revtex4-1}

\usepackage{graphicx,color}
\usepackage{amsfonts,amsmath,amsthm}

\def\B{{\cal B}}
\def\Z{{\cal Z}}
\def\M{{\cal M}}
\def\C{{\cal C}}
\def\D{{\cal D}}
\def\Y{{\cal Y}}
\def\L{{\cal L}}
\def\G{{\cal G}}
\def\Rset{\mathbb{R}}
\def\Nset{\mathbb{N}}
\def\sign{\mbox{sign}}
\def\d {\mbox{d}}

\renewcommand{\vec}[1]{\mathbf{#1}}
\newcommand{\hypgeom}[2]{\mbox{}_{#1}F_{#2}}

\def\etal{{\it et al.}}

\begin{document}

\title{Position-momentum uncertainty relations based on moments of arbitrary order}

\author{Steeve  Zozor}   \affiliation{Laboratoire  Grenoblois  d'Image,  Parole,
  Signal et Automatique (GIPSA-Lab, CNRS),  961 rue de la Houille Blanche, F-38402
  Saint Martin d'H\`eres, France} \email{steeve.zozor@gipsa-lab.grenoble-inp.fr}
\author{Mariela Portesi} \affiliation{Instituto  de F\'isica La Plata (CONICET),
  and  Departamento  de  F\'isica,  Facultad de  Ciencias  Exactas,  Universidad
  Nacional      de     La     Plata,      1900     La      Plata,     Argentina}
\email{portesi@fisica.unlp.edu.ar}
\author{Pablo  Sanchez-Moreno}  \affiliation{Instituto   Carlos  I  de  F\'isica
  Te\'orica  y  Computacional,  Universidad  de Granada,  18071-Granada,  Spain}
\affiliation{Departamento  de  Matem\'atica  Aplicada, Universidad  de  Granada,
  E-18071 Granada, Spain} \email{pablos@ugr.es}
\author{Jesus S. Dehesa} \affiliation{Instituto Carlos I de F\'isica Te\'orica y
  Computacional,     Universidad    de     Granada,     E-18071 Granada,    Spain}
\affiliation{Departamento   de   F\'isica   At\'omica,  Molecular   y   Nuclear,
  Universidad de Granada, E-18071 Granada, Spain} \email{dehesa@ugr.es}



\begin{abstract}
  The position-momentum uncertainty-like inequality based on
  moments of  arbitrary order  for $d$-dimensional quantum  systems, which  is a
  generalization  of the  celebrated Heisenberg  formulation of  the uncertainty
  principle, is improved here by  use of the  R\'enyi-entropy-based uncertainty
  relation.    The   accuracy   of   the  resulting   lower   bound   is
  physico-computationally   analyzed   for    the   two   main   prototypes   in
  $d$-dimensional physics: the hydrogenic and oscillator-like systems.
\end{abstract}

\maketitle



\section{Introduction}

The uncertainty relations play a fundamental role not only in the foundations of
quantum mechanics \cite{PriChi77,Mil90} but  also for the quantum description of
the     internal    structure     of     $d$-dimensional    physical     systems
\cite{HerAve93,LieThi02,PriChi77,Mil90,DehLop10:12}   as   well   as   for   the
development of  quantum information and  computation \cite{NieChu00,Ved06}.  The
(position-momentum) uncertainty  principle has attracted  considerable attention
since  the early  days  of  quantum mechanics  \cite{Hei27,Ken27}  up until  now
\cite{PriChi77,Mil90,FolSit97,BusHei07,BiaRud10}   because   of   its   numerous
scientific  and  technological implications.   The  first mathematical  relation
which  expresses  this  principle in  an  exact  and  quantitative form  is  the
celebrated  Heisenberg  relation  \cite{Hei27,Ken27}  which  uses  the  standard
deviation or  its square, the variance  of position and momentum,  as measure of
uncertainty; assuming $\langle \vec{x}\rangle  = \langle \vec{p}\rangle = 0$ for
notational simplicity, it reads as
\begin{equation}
\langle r^2\rangle \langle p^2\rangle \ge \frac{d^2}{4}
\label{heisenberg_relation:eq}
\end{equation}
for $d$-dimensional quantum-mechanical states.

However, this  relation is  not only too  weak but  also it is  often inadequate
\cite{Deu83,UffHil85,Uff90,BiaRud10,HilUff88}.  In  order to take  care of these
problems,  various alternative  formulations of  the uncertainty  principle have
been proposed by use of some information-theoretic uncertainty measures like the
Shannon         entropy        \cite{BiaMyc75},         R\'enyi        entropies
\cite{Bia06,ZozVig07,ZozPor08}, Tsallis entropies \cite{Raj95,PorPla96}, entropic momenta
\cite{MaaUff88}  and  Fisher  information \cite{RomSan06,SanGon06,SanPla10},  as
recently surveyed \cite{DehLop10:12,WehWin10,BiaRud10}.

Not so well  known is the moment-based uncertainty  relation developed by Angulo
\cite{Ang93,Ang94} in 1993 which can be recast \cite{DehLop10:12} under the form
\begin{equation}
\langle r^a \rangle^{\frac{2}{a}} \langle p^b \rangle^{\frac{2}{b}} \, \ge \,
\D(a,b) \, = \, \left( \frac{e \, d^{\frac{2}{a}} \,
\Gamma^{\frac{2}{d}}\left(1+\frac{d}{2} \right)}{(a e)^{\frac{2}{a}} \,
\Gamma^{\frac{2}{d}}\left(1+\frac{d}{a} \right)} \right) \, \left( \frac{e \,
d^{\frac{2}{b}} \, \Gamma^{\frac{2}{d}}\left(1+\frac{d}{2} \right)}{(b
e)^{\frac{2}{b}} \, \Gamma^{\frac{2}{d}}\left(1+\frac{d}{b} \right)} \right)
\label{HeisenbergDehesa:eq}
\end{equation}
valid for  all $(a,b)\in\Rset_+^2=(0,+\infty)^2$. This relation,  which offers a
more general and  versatile formulation of the uncertainty  principle (note that
it reduces  to the  Heisenberg inequality (\ref{heisenberg_relation:eq})  in the
particular  case  $a=b=2$), has  not  received  so  much attention  despite  the
knowledge of  the moments often  completely characterize a  probability density.
Strictly  speaking, in  the  $d$-dimensional case  and  when the  characteristic
function admits a Taylor expansion at  any order, the assertion that the moments
characterize  a distribution  is true  concerning all  the moments  of  the form
$\displaystyle \int_{\Rset^d} \prod_{i=1}^d  \left( x_i^{k_i} \rho(\vec{x}) dx_i
\right)$ for all $k_i  \in \Nset$. The assertion is no more  true when (some of)
these  moments do not  exist and/or  dealing only  with fractional  moments. For
example,  this appears  for laws  that  are not  exponentially decreasing  (e.g.
power law such as L\'evy noise).   This is known as the Hamburger moment problem
\cite[chap.  III,  \S  8]{Wid46}.   Finally,  moments of  various  orders  often
describe    fundamental   quantities    of   the    involved    quantum   system
\cite{DehLop10:12}.  Other  similar relationships  for particular values  of the
parameters  have also  been  published \cite{RomAng99,WanCar99,FolSit97}.   Note
also   that   quantities   $\langle   r^a  \rangle^{\frac{2}{a}}   \langle   p^b
\rangle^{\frac{2}{b}}$ are  insensitive to a  stretching factor in  the position
(or equivalently in  the momentum). Moreover, for specific  values of $a$ and/or
$b$, the moments are linked to physical quantities (e.g.  atomic Thomas-Fermi or
Dirac  exchanges \cite{DehLop10:12}).   Thus,  it  may offer  a  useful tool  to
quantify complexity for atomic or  chemical systems that can be complementary to
those proposed e.g.  in \cite{LopMan95,DehLop10:12,LopAng09,LopEsq10}.

In this work we deal with relation (\ref{HeisenbergDehesa:eq}) and improve it by
use  of a  R\'enyi-entropy-based approach,  in a  way similar  to  the procedure
followed by  Bialynicki-Birula and Mycielski (BBM in  short) \cite{BiaMyc75} and
Angulo     \cite{Ang93,Ang94}      used     to     obtain      the     relations
(\ref{heisenberg_relation:eq})  and  (\ref{HeisenbergDehesa:eq}),  respectively,
from the Shannon  entropy.  For this purpose we first  fix notations and briefly
review the  entropic uncertainty relations in  Section \ref{Recalls:sec}.  Then,
in Section  \ref{HeisenbergExtended:sec}, we find a  moment-based formulation of
the  uncertainty   principle  which   extends  and  generalizes   the  relations
(\ref{heisenberg_relation:eq})  and   (\ref{HeisenbergDehesa:eq}).   In  Section
\ref{Application:sec}  we  carry  out   a  computational  analysis  of  the  new
moment-based  uncertainty relation for  hydrogenic and  oscillator-like systems,
not only  because they  are the two  main quantum prototypes  in $d$-dimensional
physics  but  also  because  their  position and  momentum  moments  have  known
analytical  expressions in  terms  of  the hyperquantum  numbers  at all  orders
\cite{DehLop10:07}.    Finally,   some   conclusions   are  given   in   section
\ref{Discussions:sec}.  In the appendices we provide help to clearly discuss the
proof   of    the   moment    uncertainty   relation   described    in   Section
\ref{HeisenbergExtended:sec}.


\section{Entropic uncertainty relations: a brief review}
\label{Recalls:sec}

Let   us   denote  by   $\Psi(\vec{x})$   and  $\widehat{\Psi}(\vec{p})$   the
wavefunctions of a $d$-dimensional quantum-mechanical system in the position and
momentum spaces, respectively, so that
\[
\widehat{\Psi}(\vec{p}) = (2 \pi)^{-d/2} \int_{\Rset^d} \Psi(\vec{x}) \exp(-
\imath \vec{x}^t \vec{p}) \, \d\vec{x}
\]
where the units with $\hbar=1$ are used. The corresponding position and momentum
probability densities will be denoted as
\[
\rho(\vec{x})=|\Psi(\vec{x})|^2 \quad\text{and}\quad
\gamma(\vec{p})=|\widehat{\Psi}(\vec{p})|^2,
\]
respectively.   These   two  density  functions  are  known   to  be  completely
characterized by the knowledge of  the moments $\langle r^a\rangle$ and $\langle
p^b\rangle$   of   all   orders,   respectively,   where   $r=\|\vec{x}\|$   and
$p=\|\vec{p}\|$ denote  the Euclidean norms of the  $d$-dimensional position and
momentum single-particle operators, respectively. The position expectation value
$\langle f(r)\rangle$ is defined as
\[
\langle f(r) \rangle = \int_{\Rset^d} f(\|\vec{x}\|) \rho(\vec{x}) \, \d\vec{x},
\]
and similarly for the expectation  value $\langle f(p)\rangle$ with the momentum
density $\gamma(\vec{p})$.

For  notational simplicity,  we assume  that $\vec{x}$  and $\vec{p}$  have zero
mean, so that the variance-based  Heisenberg uncertainty relation takes the form
(\ref{heisenberg_relation:eq}). Nowadays it is well-known that there exist other
uncertainty  relations  which  are  much  more stringent.   They  are  based  on
information-theoretic quantities  such as the Shannon and  R\'enyi entropies and
the Fisher information, which provide complementary measures of the position and
momentum  probability spreading.   Let  us  here recall  the  definition of  the
R\'enyi   entropy   of   (real)    index   $\lambda\ge   0$,   $\lambda\neq   1$
\cite{Ren61,CovTho91}
\begin{equation}
H_\lambda(\rho)  = \frac{1}{1-\lambda}  \log  \int_{\Rset^d} \left(\rho(\vec{x})
\right)^\lambda  \d\vec{x}  = \frac{2  \,  \lambda}{1-\lambda} \log  \|\Psi\|_{2
  \lambda},
\label{renyi_definition_eq}
\end{equation}
which  represents an  alternative generalized  measure of  uncertainty  (lack of
information) of  a random variable  with probability density $\rho  = |\Psi|^2$.
Here, $\|\cdot\|_s$ denotes  the $L^s$ norm for functions:  $\|\Psi\|_p = \left(
  \int_{\Rset^d}  |\Psi(\vec{x})|^s   d\vec{x}  \right)^{1/s}$.   Note   that  \
$\displaystyle \lim_{\lambda \to 1} H_\lambda(\rho) = H(\rho) = - \int_{\Rset^d}
\rho(\vec{x}) \log \rho(\vec{x})  \d\vec{x}$ \ is the Shannon  entropy, that can
thus be  viewed as a special  case of the  family of R\'enyi entropies  (we will
write $H=H_1$).

To  derive an entropic  formulation of  the uncertainty  relation, the  point to
start  with is  the  Beckner relation  that links  the  $L^s$ norm  of a  (wave)
function   $\Psi(\vec{x})$  to  the   $L^q$  norm   of  its   Fourier  transform
$\widehat{\Psi}(\vec{p})$  where  $\vec{x}$  and  $\vec{p}$  are  continuous  in
$\Rset^d$, $d$ being the dimension, and  $s$ and $q$ being conjugated numbers in
the H\"older sense: $1/s  + 1/q = 1$.  This relation states  that for any $s \in
[1 ; 2]$ and $q = s / (1-s)$:
\begin{equation}
\left\| \widehat{\Psi} \right\|_q \le \left( C_{s,q} \right)^d \left\| \Psi
\right\|_s,
\label{beckner_relation_eq}
\end{equation}
where
\begin{equation}
C_{s,q} = \left( \frac{2 \pi}{s} \right)^{- \, \frac{1}{2 s}} \left( \frac{2
\pi}{q} \right)^{\frac{1}{2 q}}
\end{equation}
is the Babenko-Beckner constant.  Thus,  by taking the logarithm of the relation
(\ref{beckner_relation_eq}) with $s=2\alpha$ and $q=2\alpha^*$, one achieves the
relation \cite{Bia06,ZozVig07}
\begin{equation}
H_\alpha(\rho) + H_{\alpha^*}(\gamma) \ge d \left( \log(2 \pi) + \frac{\log(2
\alpha)}{2 (\alpha-1)} + \frac{\log(2 \alpha^*)}{2 (\alpha^*-1)} \right),
\label{RenyiEUR:eq}
\end{equation}
where   $\alpha$   and  $\alpha^*$   are   two   real   parameters  related   by
$\frac{1}{\alpha}+\frac{1}{\alpha^*}=2$, from  which we define $\alpha^*(\alpha)
=  \alpha/(2 \alpha-1)$.   In principle  $\alpha \in  \left[ \frac12  \: ;  \: 1
\right]$ but it can be seen that by symmetry (exchanging the roles of $\Psi$ and
$\widehat{\Psi}$), this relation  holds for any $\alpha \ge  1/2$.  When $\alpha
\to 1$, then  $\alpha^* \to 1$ and thus  the BBM relation \cite{BiaMyc75}
dealing with Shannon entropies is recovered
\begin{equation}
H(\rho) + H(\gamma) \ge d \left( 1 + \log(\pi) \right).
\label{ShannonEUR:eq}
\end{equation}
The     entropic    uncertainty    relations     given    in~(\ref{RenyiEUR:eq})
and~(\ref{ShannonEUR:eq}) can be recast in the more convenient product form
\begin{equation}
N_\alpha(\rho) N_{\alpha^*}(\gamma) \ge \B(\alpha),
\label{ZV:eq}
\end{equation}
with
\begin{equation}
\B(\alpha) = \frac{\alpha^{\frac{1}{\alpha-1}} \alpha^{* \:
\frac{1}{\alpha^*-1}}}{4 e^2} \quad \mbox{for } \alpha \ne 1 \qquad \mbox{and }
\qquad \B(1) = \frac{1}{4} \ ,
\label{Bound_Conj:eq}
\end{equation}
using the so-called R\'enyi $\lambda$-entropy  power
\begin{equation}
N_\lambda = \frac{1}{2 \pi e} \exp\left(\frac{2}{d} H_\lambda \right),
\end{equation}
where the limiting case $\lambda\to  1$ corresponds to the Shannon entropy power
$N  = N_1$.   BBM showed  also that  his  primary relation~(\ref{ShannonEUR:eq})
using     Shannon    entropies     {\it    does     imply}     the    Heisenberg
relation~(\ref{heisenberg_relation:eq}).   To show this,  it suffices  to search
for the maximizer  of $N(\rho)$ under variance constraint  $\langle r^2 \rangle$
fixed, that  is known to be  a Gaussian of covariance  matrix $\frac{\langle r^2
  \rangle}{d}  \,  I$  for  which   the  entropy  power  is  $\frac{\langle  r^2
  \rangle}{d}$   \cite{CovTho91,CosHer03}.    The  same   work   is  then   done
(separately) for  the momentum,  i.e.\ for $N(\gamma)$  subject to  $\langle p^2
\rangle$ fixed, to finally achieve
\begin{equation}
\langle r^2 \rangle \langle p^2 \rangle \ge d^2 N(\rho) N(\gamma)
\ge\frac{d^2}{4}
\end{equation}
and thus  the Heisenberg relation.  Heisenberg  inequality is known  to be sharp
and,  fortunately, nothing  is  lost by  this  way of  making. Indeed,  equality
between the entropy  and its maximal value  is reached if and only  if $\rho$ is
Gaussian.  Furthermore,  if (and only if)  $\rho$ is Gaussian,  $\gamma$ is also
Gaussian  with the  ``appropriate'' variance,  and thus  simultaneously,  in the
momentum space the  maximum entropy is achieved. In other words,  the sum of the
maximum entropies  corresponds to the  maximum of the sum  here. Simultaneously,
the BBM inequality becomes an equality if and only if $\rho$ is Gaussian, and
thus the succession of inequalities are equalities.

Note  now that  the relation~(\ref{ZV:eq})  with R\'enyi  entropies  given above
concerns  only  indexes  $\alpha$ and  $\alpha^*$  so  that  $2 \alpha$  and  $2
\alpha^*$       are      conjugated       in      the       H\"older      sense:
$\frac{1}{2\alpha}+\frac{1}{2\alpha^*}=1$.   Zozor {\etal}  \cite{ZozPor08} then
showed   that   the   relation~\eqref{ZV:eq}   extends  for   {\it   any}   pair
$(\alpha,\beta)$ in  $\Rset_+^2$ such that $\beta  \le \alpha^*(\alpha)$, simply
noting that  $N_\lambda$ viewed  as a function  of $\lambda$ is  decreasing (and
after decomposing  the allowed  domain for the  parameters into  three regions),
leading to
\begin{equation}
N_\alpha(\rho) N_\beta(\gamma) \ge  \Z(\alpha,\beta)
\label{ZPV:eq}
\end{equation}
where the bound is
\begin{equation}
\Z(\alpha,\beta) = \left\{\begin{array}{ll}
1/e^2 & \mbox{ for } (\alpha,\beta) \in [0 ; 1/2]^2\vspace{2mm}\\
\B(\max(\alpha,\beta)) & \mbox{ otherwise}
\end{array}\right.\label{BoundZPV:eq}
\end{equation}
with $\B$ defined in Eq.~(\ref{Bound_Conj:eq}).

Note that  on the ``conjugation curve''  $\beta = \alpha^*(\alpha)=  \alpha/ ( 2
\alpha  - 1)$,  the bound  is sharp  and  attained if  (and only  if) $\rho$  is
Gaussian,  since  it is  the  (only) case  of  equality  in the  Babenko-Beckner
relation  (see Lieb's paper~\cite{Lie90}).   Finally, let  us also  mention that
Zozor {\etal} \cite{ZozPor08} showed that  for $\beta > \alpha^*$ no uncertainty
principle exists,  in the sense  that the product  of R\'enyi entropy  powers is
just trivially non-negative.   But below the conjugation curve,  it is not known
yet neither  the sharpest  bound, nor the  states that saturate  the uncertainty
relation.


\section{The moment-based uncertainty relation}
\label{HeisenbergExtended:sec}

The  uncertainty  relation  (\ref{HeisenbergDehesa:eq})  based  on  the  moments
$\langle    r^a\rangle$    and   $\langle    p^b\rangle$    was   obtained    in
\cite{Ang93,Ang94,AngDeh92}  by use of  two elements:  the Shannon-entropy-based
BBM  relation (\ref{ShannonEUR:eq})  and  the maximizer  \cite{AngDeh92} of  the
Shannon  entropy of  the  position  (momentum) density  subject  to (s.t.)   the
constraint  $\langle r^a\rangle$  (resp. $\langle  p^b\rangle$).  Let  us remark
that such a bound cannot be  sharp.  If we denote by $\Psi_{\mathrm{max},a}$ the
wavefunction that gives  the maximizer of $N(\rho)$ s.t.\  $\langle r^a \rangle$
and  by  $\widetilde{\Psi}_{\mathrm{max},b}$  the  wavefunction  that  maximizes
$N(\gamma)$ s.t.\ $\langle p^b \rangle$, then these two functions are not linked
by  a  Fourier  transformation,  namely  $\widetilde{\Psi}_{\mathrm{max},b}  \ne
\widehat{\Psi}_{\mathrm{max},a}$, except  for the particular  case $a = b  = 2$.
Or, in other words, the sum of  the maximal entropies is not here the maximum of
the  sum.  When  deriving  the Heisenberg  relation  from the  Bialynicki-Birula
relation, although the maximization is  made separately on each Shannon entropy,
it appears that the squared roots  of the two maximizers are precisely linked by
a Fourier transformation.   Without entering into details here,  let us consider
the  example of $\rho_{\mathrm{max},a}  = \arg\max  H(\rho)$ s.t.\  $\langle r^a
\rangle$ that is a  generalized Gaussian of index $a$~\cite{Nad05,VarAaz89}. Its
squared  root is  thus a  generalized Gaussian  of index  $a/2$ and  the Fourier
transform  of  the probability  density  function  (pdf)  is not  a  generalized
Gaussian (i.e.\ not the maximizer of the other Shannon entropy): it is linked to
an $\alpha$-stable law of stability index $a/2$~\cite{ZozVig10}.

In this Section we improve the relation (\ref{HeisenbergDehesa:eq}) in a similar
way but  using the  R\'enyi entropy (\ref{renyi_definition_eq}),  which includes
the Shannon entropy as a particular case. Our procedure has the following steps:

\begin{enumerate}
\item\label{RenyiIneq:item}  Start  with  the  R\'enyi-entropy-based  inequality
  \eqref{ZPV:eq}, namely  $N_\alpha(\rho) N_\beta(\gamma) \ge \Z(\alpha,\beta)$,
  with the bound $\Z$ defined in Eq.~\eqref{BoundZPV:eq}.
\item\label{MaxRenyiR:item}  Search  for   the  maximum  R\'enyi  entropy  power
  $N_\alpha(\rho)$  s.t.\ $\langle  r^a  \rangle$.   This will  give  rise to  a
  relation   of  the   form  $\langle   r^a  \rangle^{2/a}   \ge  N_\alpha(\rho)
  \M(a,\alpha)$, where  the bound $\M$  has to be  obtained in terms of  $a$ and
  $\alpha$ (see Appendices~\ref{Maximizers:app} and~\ref{MaxEnt:app}).
\item\label{MaxRenyiP:item}  Similarly (and  separately) for  the  momentum one
  will  arrive at the  relation $\langle  p^b \rangle^{2/b}  \ge N_\beta(\gamma)
  \M(b,\beta)$
\item\label{FamilyIneq:item}    These     will    lead    to     $\langle    r^a
  \rangle^{\frac{2}{a}}  \langle  p^b  \rangle^{\frac{2}{b}} \ge  N_\alpha(\rho)
  N_\beta(\gamma)   \M(a,\alpha)   \M(b,\beta)   \ge  \M(a,\alpha)   \M(b,\beta)
  \Z(\alpha,\beta)$, for every pair $(a,b)\in\Rset_+^2$.
\item\label{MaxFamily:item} Finally,  the best bound  we can find is  $\C(a,b) =
  \max_{\alpha,\beta}  \M(a,\alpha) \M(b,\beta) \Z(\alpha,\beta)$,  where $\beta
  \le \alpha^*(\alpha)$  (other restrictions  come out that  considerably reduce
  the $(\alpha,\beta)$ domain for searching the maximum; see Appendix~C).
\end{enumerate}
It can  be shown (see Appendix~\ref{MaxBoundConj:app}) that  the desired maximum
is  {\it   on}  the  conjugation   curve  $\beta  =   \alpha^*(\alpha)$, and then
$\displaystyle \C(a,b) = \max_{\alpha} \M(a,\alpha) \M(b,\alpha^*) \B(\alpha)$.

As previously  mentioned, the bound  must be  at least the  same as the  case of
Dehesa  {\etal}   \cite{DehLop10:12},  since  the  latter   corresponds  to  the
particular situation $\alpha = \beta = 1$ in our computations.

\

The main  result of  the present effort  is summarized  here (and proved  in the
appendices):\\For any $a \ge  b > 0$ there exists an Uncertainty  Principle that can be stated
in the  following way for arbitrary-order  moments of the  position and momentum
observables in $d$-dimensional systems
\begin{equation}
\langle r^a \rangle^{\frac{2}{a}} \, \langle p^b \rangle^{\frac{2}{b}} \, \ge \,
\C(a,b) \, = \, \max_{\alpha \in D} \, \B(\alpha) \M(a,\alpha) \M(b,\alpha^*),
\label{HeisenbergMoments:eq}
\end{equation}
where     $\B(\alpha)$      is     defined     in     Eq.~\eqref{Bound_Conj:eq},
$\alpha^*=\alpha/(2\alpha-1)$,
\begin{equation}
D = \left( \max\left( \frac12 \, , \, \frac{d}{d+a} \right) \: ; \: 1 \right],
\label{Domain:eq}
\end{equation}
and the function $\M$ has the form
\begin{equation}
\M(l,\lambda) = \left\{\begin{array}{ll}
\displaystyle \!\! 2 \pi e \!\left( \!\frac{l}{\Omega B \left( \frac{d}{l} , 1 -
\frac{\lambda}{\lambda-1} - \frac{d}{l} \! \right)} \right)^{\!\!\frac{2}{d}} \!
\left( \frac{- d \, (\lambda-1)}{d (\lambda-1) + l \lambda}
\right)^{\!\!\frac{2}{l}} \left( \frac{l \lambda}{d (\lambda-1) + l \lambda}
\right)^{\!\!\frac{2}{d (\lambda-1)}}, &\vspace{1.5mm}\\
& \hspace{-2.8cm} 1 - \frac{l}{l+d} < \lambda < 1
\vspace{5mm}\\
\displaystyle \!\! 2 \pi e \! \left(\frac{l}{\Omega \, \Gamma \left( \frac{d}{l}
\right)} \right)^{\!\!\frac{2}{d}} \left( \frac{d}{l e}
\right)^{\!\!\frac{2}{l}}, & \hspace{-1cm} \lambda = 1 \vspace{5mm}\\
\displaystyle \!\! 2 \pi e \! \left(\frac{l}{\Omega \, B \left( \! \frac{d}{l} ,
\frac{\lambda}{\lambda-1} \!\right)} \right)^{\!\!\frac{2}{d}} \! \left( \frac{d
(\lambda-1)}{d (\lambda-1) + l \lambda} \right)^{\!\!\frac{2}{\lambda}} \left(
\frac{l \lambda}{d (\lambda-1)+ l \lambda} \right)^{\!\!\frac{2}{d (\lambda-1)}}, &
\hspace{-1cm} \lambda > 1
\end{array}\right.\label{HeisenbergMomentsM:eq}\end{equation}
with $\Omega=\frac{2\pi^{d/2}}{\Gamma(d/2)}$ and $B(x,y)$ the beta function.

The  case $b  \ge a  >  0$ can  be treated  using  the symmetry  (proved in  the
appendix),
\begin{equation}
\alpha_{\mathrm{opt}}(a,b) = \arg\max_{\alpha \in D} \B(\alpha) \M(a,\alpha)
\M(b,\alpha^*)
\label{alpha_opt:eq}
\end{equation}
satisfies
\begin{equation}
\alpha_{\mathrm{opt}}(b,a) = (\alpha_{\mathrm{opt}}(a,b))^*,
\label{sym_alpha_opt:eq}
\end{equation}
and then
\begin{equation}
\C(b,a) = \C(a,b).
\label{sym_C:eq}
\end{equation}
The   symmetry  on   $\alpha_{\mathrm{opt}}$  allows   also  to   conclude  that
$\alpha_{\mathrm{opt}}(a,a) =  1$ and thus  the optimal bound from  our approach
coincides   with   that   of   Angulo,   given   in~\eqref{HeisenbergDehesa:eq}.
Unfortunately, except for the case $a=b$, we have not been able yet to obtain an
analytical expression for $\C(a,b)$.

\begin{figure}[htbp]
\centerline{\includegraphics[width=16.75cm]{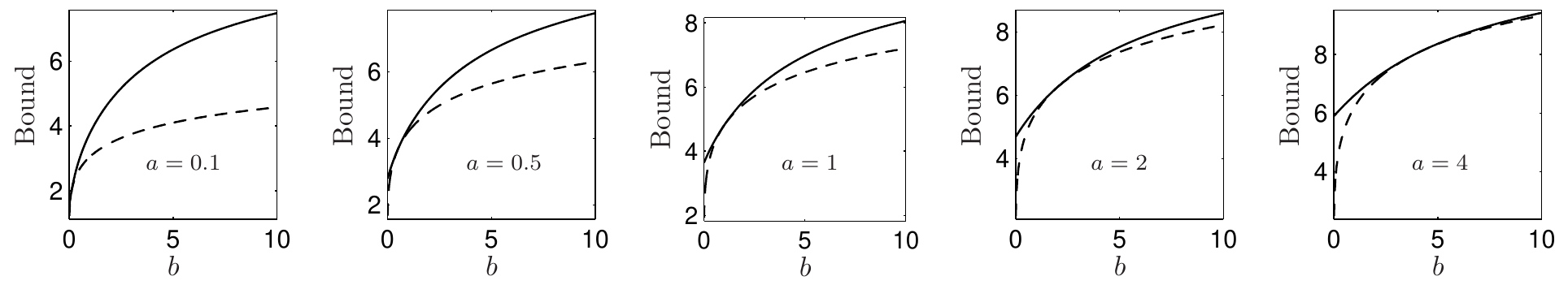}}
\caption{Bound  $\C(a,b)$  (solid  line)  given  in~(\ref{HeisenbergMoments:eq})
  compared  to $\D(a,b)$  in~\eqref{HeisenbergDehesa:eq}  (dashed line),  versus
  $b$, for  given $a=0.1,0.5,1,2$  and $4$ respectively,  with $d=5$.   For each
  value of $a$, the new bound $\C$ is always above $\D$; both functions coincide
  when $b=a$.}
\label{CotaHeisenbergG:fig}
\end{figure}

Figure~\ref{CotaHeisenbergG:fig} depicts the bound $\C(a,b)$ for given values of
$a$, as a function of $b$, compared  to the bound $\D(a,b)$.  From the figure we
see that the  bound is substantially improved when $b \ne  a$, especially as $b$
departs considerably from $a$.

\begin{figure}[htbp]
\centerline{\includegraphics[width=16.75cm]{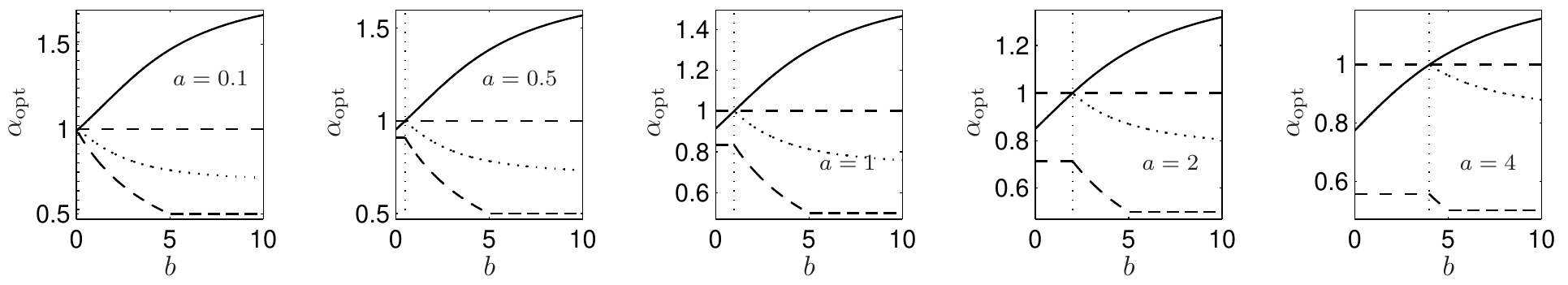}}
\caption{$\alpha_{\mathrm{opt}}(a,b)$         (solid         line)         given
  in~(\ref{alpha_opt:eq}),  versus  $b$,   for  given  $a=0.1,0.5,1,2$  and  $4$
  respectively, with $d=5$. The dotted vertical line indicates $b=a$. Thus, left
  to this line, $a \ge b >  0$ and $\alpha_{\mathrm{opt}}$ has to be searched in
  $D$ Eq.~\eqref{Domain:eq}. This  domain is indicated by the  dashed lines.  At
  the opposite,  right to  the vertical  dotted line, $b  > a$.   Thus, symmetry
  Eq.~\eqref{sym_alpha_opt:eq} is used  and $\alpha_{\mathrm{opt}}(b,a) = \left(
    \alpha_{\mathrm{opt}}(a,b) \right)^*$ is seek. Since $b>a>0$, this parameter
  is  also in  domain $D$  Eq.~\eqref{Domain:eq} (where  $b$ replaces  $a$); the
  dotted  curve represents $\alpha_{\mathrm{opt}}(b,a)$  (the solid  curve being
  $\alpha_{\mathrm{opt}}(a,b)$)  and  domain $D$  is  still  represented by  the
  dashed lines.}
\label{AlphaHeisenbergG:fig}
\end{figure}

Figure~\ref{AlphaHeisenbergG:fig}     depicts    the    optimal     $\alpha    =
\alpha_{\mathrm{opt}}$ as  a function  of $b$ in  the same configurations  as in
Fig.~\ref{CotaHeisenbergG:fig}.
The curves  illustrate that only  for $a=b$, the  optimal bound is  obtained for
$\alpha_{\mathrm{opt}} = 1$.   For $a \ne b$, a finer study  of $\M$ could allow
to even reduce the domain $D$ where $\alpha_{\mathrm{opt}}$ lies.


\section{Application  to  central  potential  problems}
\label{Application:sec}

Let    us   now   apply    and   discuss    the   minimal    uncertainty   bound
\eqref{HeisenbergMoments:eq}  for  the two  main  prototypes of  $d$-dimensional
physics:  hydrogenic and  oscillator-like systems.   But before,  let us  give a
brief review on eigensolutions for quantum systems in central potentials.


\subsection{Eigensolutions for central potentials: a brief review}

In both  hydrogenic and oscillator cases,  the quantum systems  are described by
the physical solutions of the Schr\"odinger equation
\begin{equation}
\left[ - \frac12 \nabla^2 + V(r) \right] \Psi = E \, \Psi,
\label{Schrodinger_Equation:eq}
\end{equation}
where $V(r)$  is a radial potential  and where, without loss  of generality, the
mass is set  to $m=1$.  It is well  known~\cite{YanVan99} that the wavefunctions
of a  Hamiltonian with  central potential  can be separated  out into  a radial,
$R_{E,l}(r)$, and an angular, $\Y_{\{\mu\}}(\Omega_{d-1})$, part as
\begin{equation}
\Psi_{E,\{\mu\}}(\vec{x}) = R_{E,l}(r) \Y_{\{\mu\}}(\Omega_{d-1}).
\label{Wavefunction_Radial_Angular:eq}
\end{equation}
The  position $\vec{x}  =  (x_1 ,  \ldots  , x_d)$  is  given in  hyperspherical
coordinates       as      $(r,\theta_1,\theta_2,\ldots,\theta_{d-1})      \equiv
(r,\Omega_{d-1})$, where naturally $\|\vec{x}\|  = r = \sqrt{\sum_{i=1}^d x_i^2}
\in [0  \: ;  \: +\infty)$  and $x_i =  r \left(\prod_{k=1}^{i-1}  \sin \theta_k
\right) \cos  \theta_i$ for $1 \le  i \le d$ and  with $\theta_i \in [0  \: ; \:
\pi), i < d-1$, $\theta_{d-1} \in [0  \: ; \: 2 \pi)$. By convention $\theta_d =
0$ and the empty product is the  unity.  The angular part, common to any central
potential,  is  given   by  the  hyperspherical  harmonics~\cite{YanVan99,Ave02}
$\Y_{\{\mu\}}(\Omega_{d-1})$, which are known to satisfy the eigenvalue equation
\[
\Lambda^2_{d-1} \Y_{\{\mu\}}(\Omega_{d-1}) = l \, ( l + d - 2 ) \,
\Y_{\{\mu\}}(\Omega_{d-1}),
\]
associated to the generalized angular momentum operator given by
\[
\Lambda^2_{d-1}  = -  \sum_{i=1}^{d-1} \frac{  (\sin\theta_i)^{i+1-d}  }{ \left(
    \prod_{j=1}^{i-1}         \sin\theta_j         \right)^2        }         \,
\frac{\partial}{\partial\theta_i}          \left[         (\sin\theta_i)^{d-i-1}
  \frac{\partial}{\partial\theta_i} \right].
\]
The angular  quantum numbers  $ \{\mu\}  = \{\mu_1 \equiv  l \,  ,\mu_2 \,  , \,
\ldots \, , \, \mu_{d-1} \equiv m \}$ characterize the hyperspherical harmonics,
and satisfy the  chain of inequalities $l \equiv \mu_1 \ge  \mu_2 \ge \cdots \ge
\mu_{d-2} \ge |\mu_{d-1}| \equiv |m|$.

The radial part $R_{E,l}(r)$ fulfills the second-order differential equation
\[
\left[ -\frac12 \frac{\d^2}{\d r^2} - \frac{d-1}{2 r} \frac{\d}{\d r} + \frac{ l
    (l+d-2) }{ 2 r^2} + V(r) \right] R_{E,l}(r) = E \, R_{E,l}(r),
\]
which  only depends  on  the eigenenergy  $E$,  the dimensionality  $d$ and  the
largest angular quantum number $l=\mu_1$.

Then, the quantum-mechanical position probability density for central systems is
given by
\begin{equation}
\rho_{E,\{\mu\}}(\vec{x}) = |\Psi_{E,\{\mu\}}(\vec{x})|^2 = |R_{E,l}(r)|^2 \,
|\Y_{\{\mu\}}(\Omega_{d-1})|^2.
\label{Density_Spherical:eq}
\end{equation}
It is worth remarking that this density function is normalized to unity.  Let us
bring here that
\[
\int_0^{+ \infty} r^{d-1} |R_{E,l}(r)|^2 \, \d r = 1 \quad \mbox{and} \quad
\int_{[0 ; \pi)^{d-2} \times [0 : 2 \pi)} |\Y_{\{\mu\}}(\Omega_{d-1})|^2 \,
\d\Omega_{d-1} = 1,
\]
and that the volume element can be expressed in hyperspherical  coordinates as
\[
\d\vec{x}  =  r^{d-1} \,  \d  r  \, \d\Omega_{d-1}  =  r^{d-1}  \,  \d r  \left(
  \prod_{j=1}^{d-2} (\sin\theta_j)^{d-j-1} \, \d\theta_j \right) \d\theta_{d-1}.
\]
Thus,  the   moment  $\langle   r^a\rangle$  for  the   $d$-dimensional  density
$\rho_{E,\{\mu\}}(\vec{x})$ has the expression
\begin{equation}
\langle  r^a \rangle  =  \int_0^{+\infty} r^{d+a-1} |R_{E,l}(r)|^2 \, \d r,
\label{MomentPosition:eq}
\end{equation}
which is only characterized by  the position radial wavefunction $R_{E,l}(r)$
of the particle.

From  the Fourier  transform of  $\Psi_{E,\{\mu\}}$, it  comes out  that  in the
momentum  domain the  wavefunction  $\widehat{\Psi}_{E,\{\mu\}}$ also  separates
under the form
\[
\widehat{\Psi}_{E,\{\mu\}}(\vec{p}) = M_{E,l}(p) \, \Y_{\{\mu\}}(\Omega_{d-1}).
\]
(see e.g.~\cite{Ave02,DehLop10:07,YanVan94}) with  the same hyperspherical part,
and  the radial  part  expresses  from $R_{E,l}$  through  the Hankel  transform
(e.g.~\cite{Lor54,Ave10}),
\begin{equation}
M_{E,l}(p) = p^{1-\frac{d}{2}} \int_0^{+ \infty} r^{\frac{d}{2}} R_{E,l}(r)
J_{l+\frac{d}{2}-1}(p r) \, \d r
\label{HankelRadial:eq}
\end{equation}
($J_\nu$  is  the  Bessel function  of  the  first  kind  and of  order  $\nu$).
Immediately, in  the momentum  space, the moment  $\langle p^b \rangle$  has the
expression
\begin{equation}
\langle  p^b \rangle  =  \int_0^{+\infty} p^{d+b-1} |M_{E,l}(p)|^2 \, \d p,
\label{MomentMomentum:eq}
\end{equation}
which is only characterized by  the momentum radial wavefunction $M_{E,l}(p)$ of
the particle.

These expressions have allowed to find numerous information-theoretic properties
\cite{RomSan06,SanGon06,DehLop10:07,VanYan00,Tar04,YanVan99}  of general central
potentials, particularly  the Heisenberg \cite{SanGon06}  and Fisher-information
\cite{SanGon06,RomSan06}    uncertainty   relations,   as    recently   reviewed
\cite{DehLop10:12}.


\subsection{Application to $d$-dimensional hydrogenic systems}

Let  us now  examine the  accuracy  of the  moments-based uncertainty  relations
\eqref{HeisenbergMoments:eq} for the  main prototype of $d$-dimensional systems,
namely  the hydrogenic  atom.  This  system  has been  recently investigated  in
Ref.~\cite{DehLop10:07}  in full  detail from  the information  theory  point of
view.  In this case, the potential  has the form $V(r) = - \frac{1}{r}$ (without
loss of generality, the atomic number $Z$ is taken to be 1) and the energies are
\[
E = -\frac{1}{2\eta^2},\quad \eta=n+\frac{d-3}{2},
\quad n=1, 2, \ldots
\]
where $\eta$ denotes the grand principal quantum number.  The radial part of the
eigenfunctions are thus completely calculable~\cite{DehLop10:07,Tar04,VanYan00}.
The radial wavefunction in position domain expresses as
\begin{equation}
R_{E,l}(r) = \left( \frac{\eta}{2} \right)^{\frac{d}{2}}
\sqrt{\frac{\Gamma(\eta-L)}{2\eta\Gamma(\eta+L+1)}} \:\:
\, \tilde{r}^{L-\frac{d-3}{2}} \: \exp \left( - \frac{\tilde{r}}{2} \right) \:
\L^{2L+1}_{\eta-L-1}(\tilde{r})
\label{RadialHydrogen:eq}
\end{equation}
where $L = l+\frac{d-3}{2}$, $l = 0,  \ldots , n-1$ is the grand orbital quantum
number, $\tilde{r} = \frac{2 r}{\eta}$ is a reduced (dimensionless) position and
$\L_p^q$    are   the    Laguerre   polynomials.     As   it    is    shown   in
Refs.~\cite{DehLop10:07,Tar04,VanYan00},         after         the        Hankel
transform~\eqref{HankelRadial:eq},  the radial  wavefunction in  momentum domain
expresses as
\[
M_{E,l}(p)  = 2^{2L+3}  \sqrt{\frac{\Gamma(\eta-L)}{2 \pi  \Gamma(\eta+L+1)}} \,
\Gamma(L+1)  \,  \eta^{\frac{d+1}{2}}  \frac{\tilde{p}^l}{(1+\tilde{p}^2)^{L+2}}
\:\: \G_{\eta-L-1}^{L+1}\left(\frac{1-\tilde{p}^2}{1+\tilde{p}^2}\right)
\]
where $\tilde{p}  = \eta p$ is  a reduced (dimensionless)  momentum and $\G_p^q$
are  the   Gegenbauer  polynomials.    From  these  expressions   together  with
\eqref{MomentPosition:eq}    and   \eqref{MomentMomentum:eq},   it    is   shown
\cite{DehLop10:07} that  the position and momentum moments  of arbitrary orders,
corresponding  to a  given  eigenstate characterized  by  an energy  $E$ and  an
angular  quantum  number  $l$ (or  equivalently  by  $\eta$  and $L$)  have  the
expressions:
\begin{equation}
\begin{array}{lll}
\displaystyle \langle r^a \rangle & = & \displaystyle
 \frac{\eta^{a-1}
\Gamma(2L+a+3)}{2^{a+1} \Gamma(2L+2)} \: \hypgeom{3}{2}(-\eta+L+1 , -a-1 , a+2
\: ; \: 2L+2 , 1 \: ; \: 1)\vspace{5mm}\\
\mbox{and}\vspace{5mm}\\
%
\displaystyle \langle p^b \rangle & = & \displaystyle \frac{4 \,
\Gamma(\eta+L+1) \, \Gamma\left(L+\frac{b+3}{2}\right) \,
\Gamma\left(L+\frac{5-b}{2}\right)}{\eta^{b-1} \, \Gamma(\eta-L) \,
\Gamma^2\left(L+\frac{3}{2}\right) \, \Gamma(2L+4)} \: \times\vspace{2.5mm}\\
& & \hspace{-12mm} \displaystyle \hypgeom{5}{4} \!\left( \! L\!-\!\eta\!+\!1 \,
, \, L\!+\!\eta\!+\!1 \, , \, L\!+\!1 \, , \, L\!+\frac{b\!+\!3}{2} \, , \,
L\!+\!\frac{5\!-\!b}{2} \, ; \, 2L\!+\!2 \, , \, L\!+\!\frac{3}{2} \, , \,
L\!+\!2 \, , \, L\!+\!\frac{5}{2} \, ; \, 1 \!\right)
\end{array}
\label{HydrogenMoments:eq}
\end{equation}
for $b  < 2 L  + 5$, where  $\hypgeom{p}{q}$ are the  generalized hypergeometric
functions (see e.g.~\cite[2.19.14-15]{PruBry86:v2}  and reflective properties of
hypergeometric  functions).   Note  that  the  momentum  wave  function  is  not
exponentially decreasing. The  direct consequence is that not  all moments exist
in the momentum  domain, what is reflected in the restriction  for the values of
$b$.

\begin{figure}[htbp]
\centerline{\includegraphics[width=16.75cm]{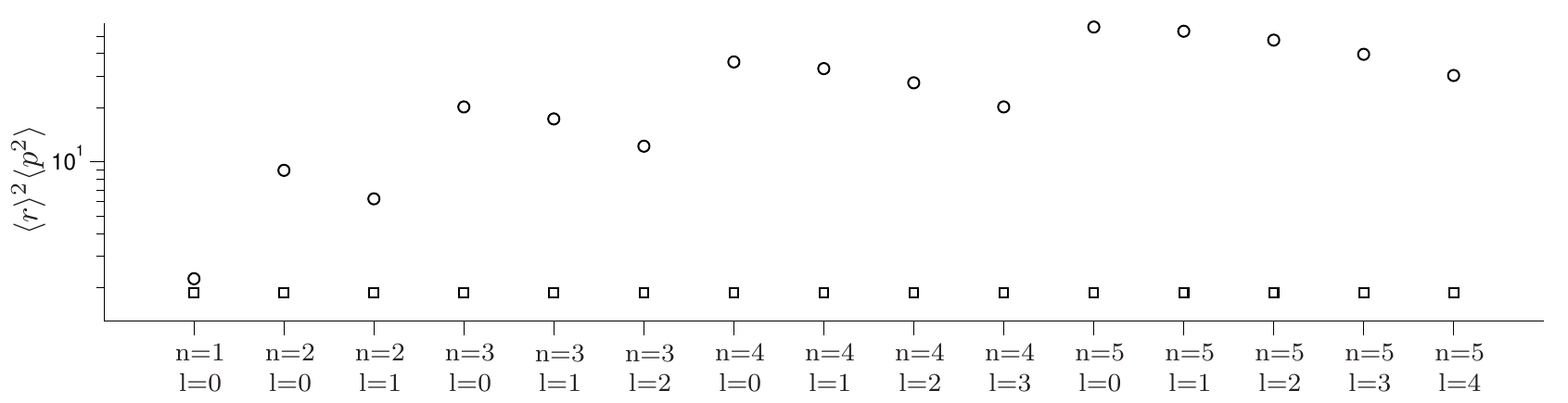}}
\caption{Product $\langle r \rangle^2  \langle p^2 \rangle$, i.e.\ $(a,b)=(1,2)$
  in eqs.~\eqref{HydrogenMoments:eq} (circles) for  the lowest energy states and
  lower                             bound                             $\C(1,2)$,
  eqs.~\eqref{HeisenbergMoments:eq}--\eqref{HeisenbergMomentsM:eq},    of   this
  product (squares), for 3-dimensional hydrogenic systems ($d=3$).}
\label{Hydrogen12:fig}
\end{figure}

\begin{figure}[htbp]
\centerline{\includegraphics[width=16.75cm]{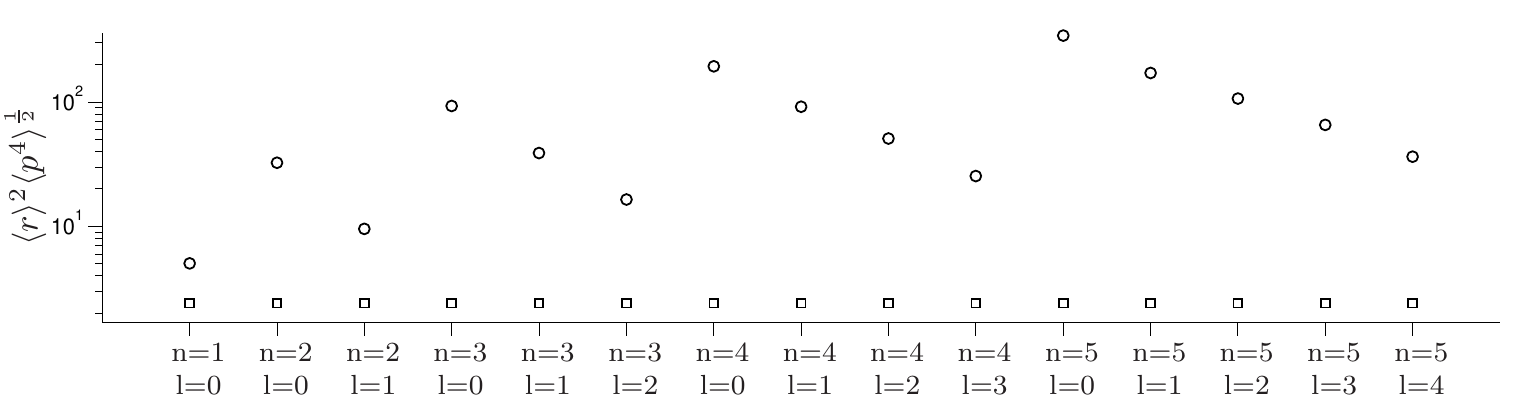}}
\caption{Product  $\langle  r \rangle^2  \langle  p^4 \rangle^{\frac12}$,  i.e.\
  $(a,b)=(1,4)$  in  eqs.~\eqref{HydrogenMoments:eq}  (circles) for  the  lowest
  energy         states        and         lower         bound        $\C(1,4)$,
  eqs.~\eqref{HeisenbergMoments:eq}--\eqref{HeisenbergMomentsM:eq},    of   this
  product (squares), for 3-dimensional hydrogenic systems ($d=3$).}
\label{Hydrogen14:fig}
\end{figure}

Thus,  the uncertainty  product $\langle  r^a \rangle^{\frac{2}{a}}  \langle p^b
\rangle^{\frac{2}{b}}$ can  be computed  and therefore studied  analytically for
hydrogenic     systems    in     $d$-dimensions.      As    an     illustration,
Fig.~\ref{Hydrogen12:fig} depicts the product $\langle r^a \rangle^{\frac{2}{a}}
\langle p^b \rangle^{\frac{2}{b}}$ computed from~\eqref{HydrogenMoments:eq}, for
$(a,b) =  (1,2)$ and  Fig.~\ref{Hydrogen14:fig} plots the  case $(a,b)  = (1,4)$
(both for  $d=3$) when  the system is  in the  state $(E,l)$, together  with the
corresponding            bound           $\C(a,b)$            given           in
\eqref{HeisenbergMoments:eq}--\eqref{HeisenbergMomentsM:eq}.

\begin{figure}[htbp]
\centerline{\includegraphics[width=16.75cm]{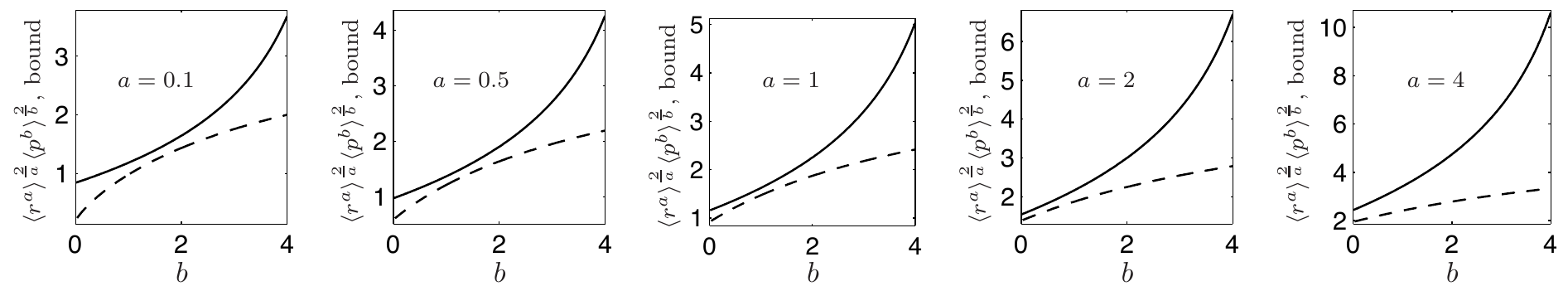}}
\caption{Product     $\langle    r^a    \rangle^{\frac{2}{a}}     \langle    p^b
  \rangle^{\frac{2}{b}}$ (solid  lines) in the ground state  ($n=1$, $l=0$), for
  fixed $a=0.1$, $a=0.5$,  $a=1$, $a=2$ and $a=4$ respectively,  and lower bound
  (dashed lines) for 3-dimensional hydrogenic systems ($d=3$).}
\label{HydrogenA:fig}
\end{figure}

We can  see from both figures that,  although not sharp, the  bound $\C(a,b)$ is
close   to  the   product   $\langle  r^a   \rangle^{\frac{2}{a}}  \langle   p^b
\rangle^{\frac{2}{b}}$ for the ground state ($n = 1$ and $l = 0$). However, when
$n$ increases, the discrepancy from  the bound increases (and decreases with $l$
for fixed  $n$). The same behavior  occurs for other pairs  $(a,b)$ whatever the
dimensionality  $d$. Since  hydrogenic systems  belong to  the family  of radial
potential systems, this  suggests that a refinement can be  found in the context
of  radial systems  as  already  done for  the  usual variance-based  Heisenberg
inequality, and  for Fisher information-based versions~\cite{RomSan06,SanGon06}.
To give  a further  illustration, Fig.~\ref{HydrogenA:fig} depicts  $\langle r^a
\rangle^{\frac{2}{a}} \langle  p^b \rangle^{\frac{2}{b}}$ as a  function of $b$,
for fixed $a$, and  for the ground state ($n=1$, $l=0$) in  3 dimensions. In all
the cases shown, we observe the existence  of a value of $b$ that is ``optimum''
in the sense  that the uncertainty product is close to  the bound proposed here,
corresponding to a  situation of low generalized uncertainty.   As $b$ increases
(up to  $2 L +  5 = 5$  for the ground state  in 3 dimensions),  the uncertainty
departs from our bound. Finally, one  observes for the tested values of $a$ that
the lower  bound has a concave  behavior versus $b$, while  the product $\langle
r^a \rangle^{\frac{2}{a}}  \langle p^b \rangle^{\frac{2}{b}}$  exhibits a convex
behavior. This  suggests the existence of  an optimal value of  $b$ (function of
$a$) in terms of low discrepancy from the bound.

\subsection{Application to $d$-dimensional oscillator-like systems}

Let us consider now  a potential of the form $V(r) =  \frac12 r^2$ (without loss
of generality, the product mass squared pulsation is taken as $m \omega^2 = 1$).
In this case, the energies are
\[
E = 2 n + l + \frac{d}{2}, \quad n=0, 1, \ldots \quad \mbox{and} \quad l=0 , 1 ,
\ldots
\]
and    the   radial    parts   of    the   wavefunctions    are    again   known
\cite{LouSha60:p2}. They express as
\begin{equation}
R_{E,l}(r) = \sqrt{\frac{2\Gamma(n+1)}{\Gamma(n+l+d/2)}} \:\: r^l \: \exp \left(
- \frac{r^2}{2} \right) \: \L^{l+d/2-1}_{n}(r^2)
\label{RadialOscillator:eq}
\end{equation}
and    $M_{E,,l}(p)   =    R_{E,l}(p)$.    Comparing~\eqref{RadialOscillator:eq}
with~\eqref{RadialHydrogen:eq}, after  a change of variables  $\tilde{r} = r^2$,
one can easily show from \eqref{HydrogenMoments:eq} that the statistical moments
write down as
\begin{equation}
\begin{array}{lll}
\displaystyle \langle r^a \rangle & = & \displaystyle
\frac{\Gamma\left(l+\frac{d+a}{2}\right)}{\Gamma\left(l+\frac{d}{2}\right)} \:
\hypgeom{3}{2}\left(-n , -\frac{a}{2} , \frac{a}{2}+1 \: ; \: L+\frac{d}{2} , 1
\: ; \: 1 \right)\vspace{5mm}\\
\mbox{and}\vspace{5mm}\\
\langle p^b \rangle & = & \displaystyle
\frac{\Gamma\left(l+\frac{d+b}{2}\right)}{\Gamma\left(l+\frac{d}{2}\right)} \:
\hypgeom{3}{2}\left(-n , -\frac{b}{2} , \frac{b}{2}+1 \: ; \: L+\frac{d}{2} , 1
\: ; \: 1 \right)
\end{array}
\label{OscillatorMoments:eq}
\end{equation}
(see also Ref.~\cite{LouSha60:p3} for special cases).

\begin{figure}[htbp]
\centerline{\includegraphics[width=16.75cm]{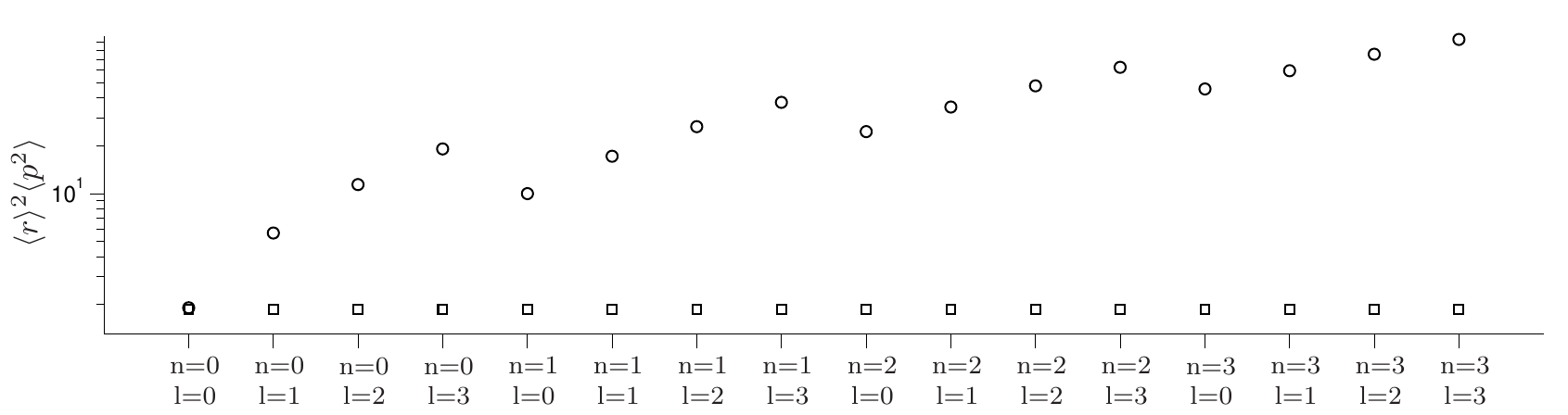}}
\caption{Product $\langle r \rangle^2  \langle p^2 \rangle$, i.e.\ $(a,b)=(1,2)$
  in eqs.~\eqref{HydrogenMoments:eq} (circles) for  the lowest energy states and
  lower                             bound                             $\C(1,2)$,
  eqs.~\eqref{HeisenbergMoments:eq}--\eqref{HeisenbergMomentsM:eq},    of   this
  product (squares) for 3-dimensional harmonic oscillators ($d=3$).}
\label{Oscillator12:fig}
\end{figure}

\begin{figure}[htbp]
\centerline{\includegraphics[width=16.75cm]{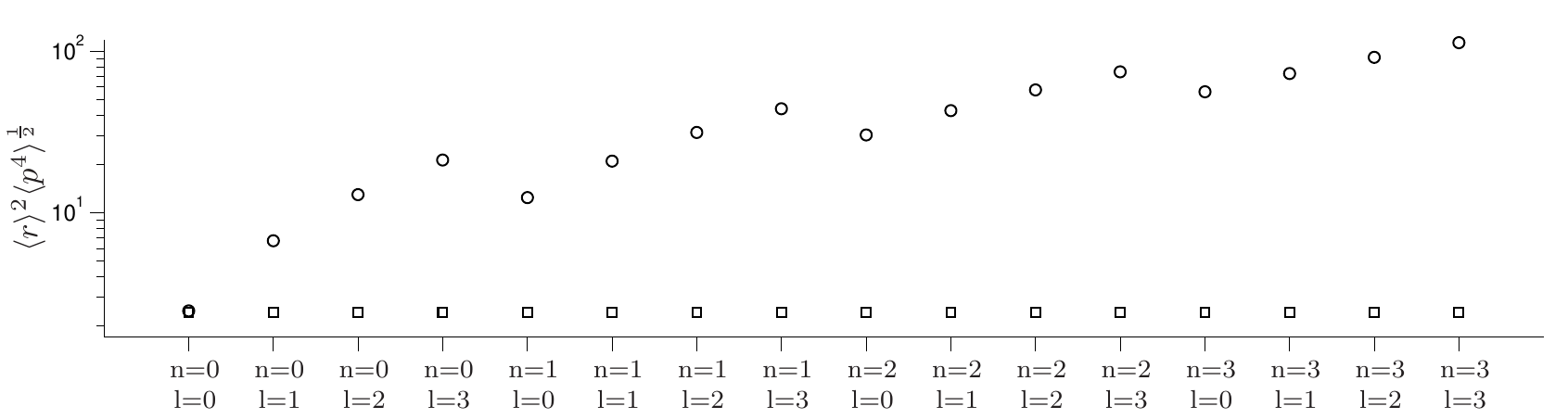}}
\caption{Product  $\langle  r \rangle^2  \langle  p^4 \rangle^{\frac12}$,  i.e.\
  $(a,b)=(1,4)$  in  eqs.~\eqref{HydrogenMoments:eq}  (circles) for  the  lowest
  energy         states        and         lower         bound        $\C(1,4)$,
  eqs.~\eqref{HeisenbergMoments:eq}--\eqref{HeisenbergMomentsM:eq},    of   this
  product (squares) for 3-dimensional harmonic oscillators ($d=3$).}
\label{Oscillator14:fig}
\end{figure}

Figure~\ref{Oscillator12:fig}  describes   the  moments  product   $\langle  r^a
\rangle^{\frac{2}{a}}      \langle     p^b      \rangle^{\frac{2}{b}}$     using
\eqref{OscillatorMoments:eq},       for      $(a,b)      =       (1,2)$      and
Fig.~\ref{Oscillator14:fig} exhibits the case  $(a,b) = (1,4)$ (both for $d=3$),
together     with    the    corresponding     bound    $\C(a,b)$     given    in
\eqref{HeisenbergMoments:eq}--\eqref{HeisenbergMomentsM:eq}.

We can see from  these figures also that even if not  sharp, the bound $\C(a,b)$
is  very close  to the  product $\langle  r^a \rangle^{\frac{2}{a}}  \langle p^b
\rangle^{\frac{2}{b}}$, for the  ground state ($n = 0$ and $l  = 0$). The global
behavior  is similar  to what  happens for  the hydrogenic  systems: there  is a
discrepancy from the bound as $n$ increases. But here, the discrepancy increases
also  with $l$ when  $n$ is  fixed.  The  same behavior  occurs for  other pairs
$(a,b)$ and  whatever the  dimension $d$.  In  fact, when observing  more finely
$\langle r \rangle^2  \langle p^2 \rangle$ and $\langle  r \rangle^2 \langle p^4
\rangle^{\frac12}$,  it appears that  these products  essentially depend  on the
energy level, i.e the  values of these products for a fixed  value of $2n+l$ are
very close  (see e.g.  $n=0,  l=2$ or  $n=1, l=0$). This  was true also  for the
hydrogenic  systems,  but  it  is  strongly more  pronounced  for  the  harmonic
oscillator. All these observations  reinforce our ``conjecture'' that refinement
can be  found in the context of  radial systems, for moments'  orders other than
$a=b=2$, at least in terms of energy levels.

\begin{figure}[htbp]
\centerline{\includegraphics[width=16.75cm]{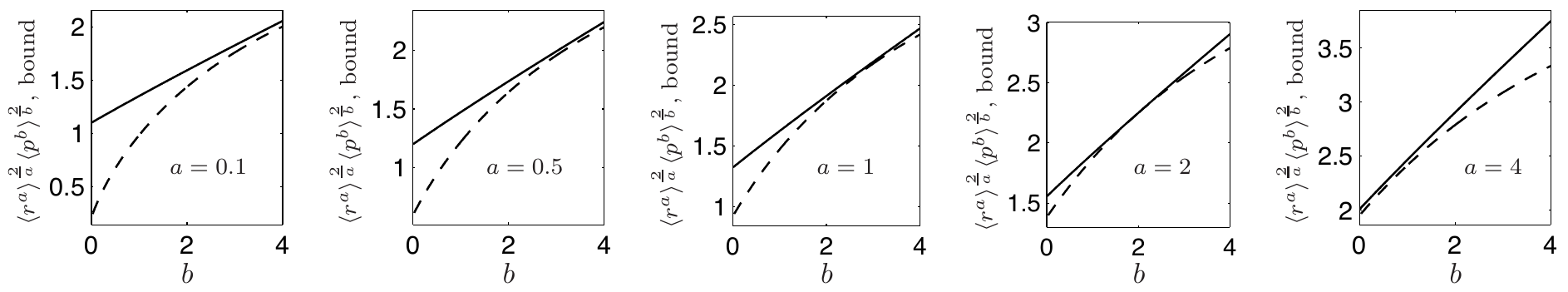}}
\caption{Product     $\langle    r^a    \rangle^{\frac{2}{a}}     \langle    p^b
  \rangle^{\frac{2}{b}}$ (solid  lines) in the ground state  ($n=0$, $l=0$), for
  fixed $a=0.1$, $a=0.5$,  $a=1$, $a=2$ and $a=4$ respectively,  and lower bound
  (dashed lines) for 3-dimensional oscillators ($d=3$).}
\label{OscillatorA:fig}
\end{figure}

A further illustration is given by Fig.~\ref{OscillatorA:fig} where $\langle r^a
\rangle^{\frac{2}{a}} \langle p^b \rangle^{\frac{2}{b}}$ versus $b$ is depicted,
for  different fixed values  of $a$,  in the  case of  the ground  state ($n=0$,
$l=0$).

Globally the  behavior of  the moments' product  compared to the  bound observed
here is  similar to  that of the  hydrogenic systems.  However,  the discrepancy
from the  bound is less  pronounced for the  harmonic oscillator (in  the ground
state) than  for the hydrogen systems.  Note  that the bound is  achieved in the
case  when  $a=b=2$.  This  case  corresponds  to  the classical  variance-based
Heisenberg inequality.   Moreover, the ground  state of the oscillator  leads to
the  Gaussian  pdf  $\rho$  (and  $\gamma$): in  this  case  the  variance-based
Heisenberg inequality is saturated.  One  can again observe the convexity of the
product  $\langle r^a  \rangle^{\frac{2}{a}} \langle  p^b \rangle^{\frac{2}{b}}$
(in  fact almost  linear): together  with the  observed concavity  of  the lower
bound, this  reinforce our conjecture  on the existence  of an optimal  value of
$b(a)$ in  terms of low discrepancy from  the bound. This remains  to be studied
more systematically and more deeply.


\section{Discussion and conclusions}
\label{Discussions:sec}

In  this  paper  we  have  proposed  an improved  version  of  the  moment-based
mathematical  formulation of  the position--momentum  uncertainty  principle for
quantum  systems  that generalizes  the  seminal  variance-based formulation  of
Heisenberg.   The  main  result  of  this  contribution  is  formalized  in  Eq.
\eqref{HeisenbergMoments:eq}    together     with    eqs.     \eqref{Domain:eq},
\eqref{HeisenbergMomentsM:eq}   and   \eqref{Bound_Conj:eq}:   $   \langle   r^a
\rangle^{\frac{2}{a}}  \langle  p^b   \rangle^{\frac{2}{b}}  \ge  \C(a,b)$.   In
contrast to  the entropic uncertainty relations (like  Eq.  \eqref{ZPV:eq}), the
new  formulation  is  based   on  spreading  measures  which  describe  physical
observables. Our present approach suffers, however, from the fact that the lower
bound $\C(a,b)$ found here for the  product of the position and momentum moments
for arbitrary  $a$ and  $b$ is not  sharp.  To  tackle this issue  a variational
approach may be envisaged, although it is a difficult task.  Another alternative
might be to  employ appropriate Sobolev-like inequalities, as  done for entropic
formulations (see e.g.~Ref.~\cite{BiaMyc75,Bia06,ZozVig07,Raj95}).

The new moment-based uncertainty relation is physico-computationally analyzed in
some $d$-dimensional  quantum systems. Precisely, the bound  of the moment-based
uncertainty relation  is compared to the  product of the  moments for hydrogenic
and oscillator-like systems.  In both  cases analytic expressions of the moments
exist in terms of  hypergeometric functions (eqs. \eqref{HydrogenMoments:eq} and
\eqref{OscillatorMoments:eq}  respectively).   Our   results  suggest  that  the
improvement  of this  relation  for general  central  potentials seems  possible
whatever the  orders $a$ and  $b$ of  the moments, at  least in terms  of energy
levels.    Such   an   improvement   exists  in   the   variance-based   context
$a=b=2$~\cite{RomSan06,SanGon06}, but for moments  of arbitrary order this issue
is  a fully  open problem  which deserves  to be  variationally solved  for both
fundamental and applied  reasons.  This suggests also that  the product $\langle
r^a \rangle^{\frac{2}{a}} \langle p^b \rangle^{\frac{2}{b}}$ can be envisaged as
a  useful  tool to  quantify  the  complexity and organization  of various  physical
systems.  However,  the properties  of such a  complexity measure should be analyzed in more detail.


\begin{acknowledgements}
S.Z.\ and  M.P.\ are  grateful  to  the  CNRS (France)  and  CONICET
(Argentina)  for  the  cooperation  grant  that  partially  enabled  this  work.
M.P.\  also   acknowledges   financial    support   by   CONICET   and   ANPCyT.
P.S.-M.\ and J.S.D.\ are very grateful to Junta de Andaluc\'ia
(Spain) for  the grants FQM-2445 and  FQM-4643, and the Ministerio  de Ciencia e
Innovaci\'on (Spain) for the grant FIS2008-02380.
\end{acknowledgements}

\appendix


\section{Evaluation of the maximizers of  the R\'enyi entropy power under moment
  constraint}
\label{Maximizers:app}

In   section   \ref{HeisenbergExtended:sec}   (steps  \ref{MaxRenyiR:item}   and
\ref{MaxRenyiP:item}) we established the  necessity of searching for the maximum
of the R\'enyi entropy power $N_\alpha(\rho)$ subjected to given moment $\langle
r^a \rangle$ and $N_\beta(\gamma)$  s.t. $\langle p^b \rangle$. This variational
problem   has   been   tackled   and   partially  solved   by   Dehesa   {\etal}
\cite{DehGal89}. Similarly to what is  done for variance constraint, the problem
is  to  maximize the  frequency  entropic  moment of  order  $\lambda  > 0$,  an
increasing  function   of  the  entropy   power,  $\displaystyle  \int_{\Rset^d}
f(\vec{x})^\lambda  \d\vec{x}$  s.t.\  $\displaystyle \int_{\Rset^d}  f(\vec{x})
\d\vec{x}  =  1$  and  $\displaystyle  \int_{\Rset^d}  \|\vec{x}\|^l  f(\vec{x})
\d\vec{x} = \langle r^l \rangle$, with  $l > 0$ and where $\|\vec{x}\|=r$ is the
Euclidean norm of $\vec{x}$. Note that we work here with the variables $\vec{x}$
and $r$, but the results obtained will be valid in the momentum domain, changing
$\vec{x}$ to $\vec{p}$ and $r$ to $p$.  Then, we have to maximize $\displaystyle
\int_{\Rset^d} \left(  f(\vec{x})^\lambda -  \mu f(\vec{x}) -  \nu \|\vec{x}\|^l
  f(\vec{x}) \right) \d\vec{x}$, where $\mu$ and $\nu$ are the Lagrange factors.
From the corresponding Euler--Lagrange equation, one obtains that $f$ must be of
the  form   $f(\vec{x})  =   \left(  \frac{\mu  +   \nu  \|\vec{x}\|^l}{\lambda}
\right)_+^{\frac{1}{\lambda-1}}$, where $(y)_+ = \max(y,0)$.  With integrability
arguments ($f$ must be  a pdf, and thus positive and integrable),  $\mu > 0$ and
$\nu$ must  have the sign  of $1-\lambda$, and  thus the pdf that  maximizes the
entropy power  $N_\lambda$ s.t.  $\langle r^l  \rangle$ can be  recast under the
form
\begin{equation}
f_{\lambda,l}(\vec{x}) = C \left( 1 - (\lambda-1)
(\|\vec{x}\|/\delta)^l\right)_+^{\frac{1}{\lambda-1}}.
\label{gene_q_gaussian:eq}
\end{equation}
This  pdf is  sometimes called  generalized  Gaussian~\cite{TsaMen98,And09}, but
this terminology is  not adequate. Indeed, when $\lambda \to  1$, this pdf tends
to  $f_{1,l}(\vec{x}) =  C \exp(-\|\vec{x}/\delta\|^l)$  that is  also sometimes
named  generalized Gaussian  (or  also Kotz-type)~\cite{VarAaz89,KotNad01,Kan94}
(and   also   sometimes   as   stretched  exponential   or   power   exponential
\cite{Kan94,Koz02}).   Furthermore,   when  $l  =   2$  one  can   recognize  in
\eqref{gene_q_gaussian:eq} the well known  $q$-Gaussian (also known as Student-t
or Student-r  depending on the sign  of $1-\lambda$), where $q  = 2-\lambda$ and
thus   the  generalization   \eqref{gene_q_gaussian:eq}  is   known   under  the
terminology of {\em  stretched $q$-exponential}~\cite{DehGal89,TsaLen02} or even
{\em  generalized  $q$-Gaussian}  of  parameter $q=2-\lambda$  and  (stretching)
parameter $l$.

Constants $C$  and $\delta$  are to  be determined so  that the  constraints are
satisfied. The normalization constraint reads
\begin{eqnarray*}
1 & = & C \, \int_{\Rset^d} \left( 1 - (\lambda-1)
(\|\vec{x}\|/\delta)^l\right)_+^{\frac{1}{\lambda-1}} \, \d\vec{x}\\
& = & C \, \Omega \, \int_0^{+\infty} r^{d-1} \left( 1 - (\lambda-1)
(r/\delta)^l\right)_+^{\frac{1}{\lambda-1}} \, \d r\\
& = & \frac{C \, \Omega \delta^d}{l |\lambda-1|^{d/l}} \, \int_0^{+\infty}
t^{d/l-1} \left( 1 - \sign(\lambda-1) \, t\right)_+^{\frac{1}{\lambda-1}} \, \d t
\end{eqnarray*}
where  the second line  comes from~\cite[Eq.   4.642]{GraRyz07}, with  $\Omega =
\frac{2 \pi^{d/2}}{\Gamma(d/2)}$ that is the surface of the $d$-dimensional unit
sphere and  where the  third line  comes from the  change of  variable $r=\delta
(t/|1-\lambda|)^{1/l}$.     The   integral   term,    that   we    will   denote
$B_1(l,\lambda)$, expresses via the  beta function $B(x,y) = \Gamma(x) \Gamma(y)
/  \Gamma(x+y)$,  from~\cite[Eq.   8.380-1  \& 8.380-3]{GraRyz07},  one  finally
obtains
\begin{equation}
1 = \frac{C \, \Omega \delta^d}{l \, |\lambda-1|^{d/l}} B_1(l,\lambda)
\label{constraint_pdf:eq}
\end{equation}
where
\begin{equation}
B_1(l,\lambda) = \left\{\begin{array}{lll}
B \left( \frac{d}{l} \: , \: \frac{\lambda}{\lambda-1} \right) & \mbox{ if } &
\lambda > 1 \vspace{2.5mm}\\
B \left( \frac{d}{l} \: , \: 1 + \frac{\lambda}{1-\lambda} - \frac{d}{l} \right)
& \mbox{ if } & 1 - \frac{l}{d} < \lambda < 1
\end{array}\right.
\label{beta_function_pdf:eq}
\end{equation}
Indeed, the integral converges provided that $\lambda > 1 - l/d$.

In the same vein, the power moment constraint writes
\begin{eqnarray*}
\langle r^l \rangle & = & C \, \int_{\Rset^d} \|\vec{x}\|^l \left( 1 -
(\lambda-1) (\|\vec{x}\|/\delta)^l\right)_+^{\frac{1}{\lambda-1}} \, \d\vec{x}\\
& = & C \, \Omega \, \int_0^{+\infty} r^{l+d-1} \left( 1 - (\lambda-1)
(r/\delta)^l\right)_+^{\frac{1}{\lambda-1}} \, \d r\\
& = & \frac{C \, \Omega \delta^{l+d}}{l |\lambda-1|^{d/l+1}} \, \int_0^{+\infty}
t^{d/l} \left( 1 - \sign(\lambda-1) \, t\right)_+^{\frac{1}{\lambda-1}} \, \d t
\end{eqnarray*}
where the integral  term, denoted here $B_m(l,\lambda)$, expresses  via the beta
function from~\cite[Eq. 8.380-1 \& 8.380-3]{GraRyz07}, leading to
\begin{equation}
\langle r^l \rangle = \frac{C \, \Omega \delta^{l+d}}{l |\lambda-1|^{d/l+1}}
B_m(l,\lambda)
\label{constraint_moment:eq}
\end{equation}
where
\begin{equation}
B_m(l,\lambda) = \left\{\begin{array}{lll}
B \left( \frac{d}{l} +1 \: , \: \frac{\lambda}{\lambda-1} \right) & \mbox{ if }
& \lambda > 1 \vspace{2.5mm}\\
B \left( \frac{d}{l} +1 \: , \: \frac{\lambda}{1-\lambda} - \frac{d}{l} \right)
& \mbox{ if } & 1 - \frac{l}{d+l} < \lambda < 1
\end{array}\right.
\label{beta_function_moment:eq}
\end{equation}
Note that the existence of the latter integral implies a stronger restriction to
$\lambda$ than the one coming from the normalization, that is we require now:
\begin{equation}
\lambda > 1-\frac{l}{d+l} = \frac{d}{d+l}.
\label{domain_lambda:eq}
\end{equation}

In both constraints, the case $\lambda = 1$ can be recovered by letting $\lambda
\to   1^+$    or   $\lambda   \to   1^-$    (from   \cite[6.1.47]{AbrSte70}   or
\cite[8.328-1]{GraRyz07}, $\lim_{y  \to \infty} B(x,y) x^y =  \Gamma(x)$): it is
not needed to treat this case separately.


\section{Maximal entropy power $N_\lambda$ and bound for the moment $\langle r^l
  \rangle$}
\label{MaxEnt:app}

Following  the procedure  proposed in  section  \ref{HeisenbergExtended:sec}, we
discuss here the  bounds for the moments $\langle r^a  \rangle$ and $\langle p^b
\rangle$.    From~\eqref{gene_q_gaussian:eq},  the  maximal   $\lambda$-norm  of
$f_{\lambda,l}(\vec{x}$ to the power $\lambda$ takes the form
\begin{eqnarray*}
\|f_{\lambda,l}\|_\lambda^\lambda & = & C^\lambda \, \int_{\Rset^d} \left( 1 -
(\lambda-1) (\|\vec{x}\|/\delta)^l\right)_+^{\frac{\lambda}{\lambda-1}} \,
\d\vec{x}\\
& = & C^\lambda \, \Omega \, \int_0^{+\infty} r^{d-1} \left( 1 - (\lambda-1)
(r/\delta)^l\right)_+^{\frac{\lambda}{\lambda-1}} \, \d r\\
& = & \frac{C^\lambda \, \Omega \delta^d}{l |\lambda-1|^{d/l}} \,
\int_0^{+\infty} t^{d/l-1} \left( 1 - \sign(\lambda-1) \,
t\right)_+^{\frac{\lambda}{\lambda-1}} \, \d t.
\end{eqnarray*}
Then, from~\cite[Eq.  8.380-1 \& 8.380-3]{GraRyz07} we obtain
\begin{equation}
\|f_{\lambda,l}\|_\lambda^\lambda = \frac{C^\lambda \, \Omega \delta^d}{l
|\lambda-1|^{d/l}} B_h(l,\lambda),
\end{equation}
where we have defined
\begin{equation}
B_h(l,\lambda) = \left\{\begin{array}{lll}
B \left( \frac{d}{l} \: , \: \frac{\lambda}{\lambda-1} + 1 \right) & \mbox{ if }
& \lambda > 1 \vspace{2.5mm}\\
B \left( \frac{d}{l} \: , \: \frac{\lambda}{1-\lambda} - \frac{d}{l} \right) &
\mbox{ if } & 1 - \frac{l}{d+l} < \lambda < 1
\end{array}\right.
\label{beta_function_entropy:eq}
\end{equation}
that adds no new restriction on $\lambda$.
Thus, the maximal value of the R\'enyi entropy power is
\begin{eqnarray*}
N_\lambda(f_{\lambda,l}) & = & \frac{1}{2 \pi e}
\left(\|f\|_\lambda^{\frac{\lambda}{1-\lambda}}\right)^{\frac{2}{d}}\\
& = & \frac{1}{2 \pi e} \left(C^{\frac{\lambda}{1-\lambda}} \left( \frac{\Omega
\delta^d}{l |\lambda-1|^{d/l}} \right)^{\frac{1}{1-\lambda}}
B_h^{\frac{1}{1-\lambda}}\right)^{\frac{2}{d}}\\
& = & \frac{1}{2 \pi e} \left(C^{-1} \left( \frac{C \Omega \delta^d}{l
|\lambda-1|^{d/l}} B_1 \right)^{\frac{1}{1-\lambda}}
\left(\frac{B_h}{B_1}\right)^{\frac{1}{1-\lambda}}\right)^{\frac{2}{d}}\\
& = & \frac{1}{2 \pi e} \left(C^{-1}
\left(\frac{B_h}{B_1}\right)^{\frac{1}{1-\lambda}}\right)^{\frac{2}{d}}\\
\end{eqnarray*}
from~\eqref{constraint_pdf:eq} and  where the arguments  of $B_1$ and  $B_h$ are
omitted    for    simplicity.     Taking    the   ratio    $\frac{\langle    r^l
  \rangle^{d/l}}{1^{d/l+1}}$          from~\eqref{constraint_pdf:eq}         and
\eqref{constraint_moment:eq}, one obtains
\begin{eqnarray*}
C^{-1} = \frac{\Omega}{l} B_1 \left(\frac{B_1}{B_m}\right)^{d/l} \langle
r^l\rangle^{d/l},
\end{eqnarray*}
that gives
\begin{eqnarray*}
N_\lambda(f_{\lambda,l}) = \frac{1}{2 \pi e} \left( \frac{\Omega B_1}{l}
\left(\frac{B_1}{B_m}\right)^{d/l}
\left(\frac{B_h}{B_1}\right)^{\frac{1}{1-\lambda}} \langle
r^l\rangle^{d/l}\right)^{\frac{2}{d}}.
\end{eqnarray*}
One can simplify a little bit this expression by considering the parameter
\begin{equation}
\mu = \mu(\lambda) = \frac{\lambda}{\lambda-1}
\label{mu:eq}
\end{equation}
that governs the maximal entropy power, with  $\mu > 1$ or $\mu < - d/l$. Noting
that
\begin{equation}
\frac{B_1}{B_m} = \sign(\mu) \frac{d + l \mu}{d}
\hspace{1cm} \mbox{and}\hspace{1cm} \frac{B_h}{B_1} = \frac{l \mu}{d + l \mu}
\end{equation}
so that
\begin{equation}
N_\lambda(f_{\lambda,l}) = \frac{1}{2 \pi e} \left( \frac{\Omega
B_1(l,\lambda)}{l} \left( \frac{d + l \mu}{\sign(\mu) \, d}\right)^{\frac{d}{l}}
\left( \frac{d + l \mu}{l \mu}\right)^{\mu-1} \langle
r^l\rangle^{d/l}\right)^{\frac{2}{d}}.
\end{equation}
We  finally  obtain   that  the  R\'enyi  entropy  power   of  any  pdf  $\rho$,
$N_\lambda(\rho)            =            \frac{1}{2            \pi            e}
\left(\|\rho\|_\lambda^{\frac{\lambda}{1-\lambda}} \right)^{\frac{2}{d}}$, s.t.\
fixed  $\langle  r^l  \rangle$, is  bounded  from  above  by the  maximum  value
$N_\lambda(f_{\lambda,l})$. Therefore we can write
\begin{equation}
\langle r^l\rangle^{2/l} \ge N_\lambda(\rho)  \M(l,\lambda)
\label{BoundMoment:eq}
\end{equation}
where function $\M$ expresses as
\begin{equation}
\M(l,\lambda) = \left\{\begin{array}{lll} \displaystyle 2 \pi e
\left(\frac{l}{\Omega \, B \left( \frac{d}{l} , \mu \right)}
\right)^{\frac{2}{d}} \left( \frac{d}{d + l \mu} \right)^{\frac{2}{l}} \left(
\frac{l \mu}{d + l \mu} \right)^{\frac{2 (\mu-1)}{d}} & \mbox{ if } & \lambda >
1\vspace{5mm}\\
\displaystyle 2 \pi e \left(\frac{l}{\Omega \, \Gamma \left( \frac{d}{l}
\right)} \right)^{\frac{2}{d}} \left( \frac{d}{l e} \right)^{\frac{2}{l}} &
\mbox{ if } & \lambda = 1 \vspace{5mm}\\
\displaystyle 2 \pi e \left(\frac{l}{\Omega \, B \left( \frac{d}{l} , 1 - \mu -
\frac{d}{l} \right)} \right)^{\frac{2}{d}} \left( - \, \frac{d}{d + l \mu}
\right)^{\frac{2}{l}} \left( \frac{l \mu}{d + l \mu} \right)^{\frac{2
(\mu-1)}{d}} & \mbox{ if } & 1 - \frac{l}{l+d} < \lambda < 1
\end{array}\right.\label{BoundM:eq}\end{equation}
with
\begin{equation}
\mu = \frac{\lambda}{\lambda-1}
\end{equation}
and where $\M(l,1)  = \lim_{\lambda \to 1} \M(l,\lambda)$  from the first and/or
second expression of $\M$ and~\cite[6.1.41]{AbrSte70}.


\section{Generalized Heisenberg-like uncertainty relation}
\label{GeneHeisenberg:app}

Using~\eqref{BoundMoment:eq} applied  to $r$ with  $l=a$ and $\lambda  = \alpha$
and  applied to  $p$  with  $l=b$ and  $\lambda=\beta$  respectively, and  using
\eqref{ZPV:eq},    we    achieve    the    relation   established    in    point
\ref{FamilyIneq:item} of section \ref{HeisenbergExtended:sec}
\begin{equation}
\langle r^a\rangle^{\frac{2}{a}} \, \langle p^b \rangle^{\frac{2}{b}} \ge
\Z(\alpha,\beta) \M(a,\alpha) \M(b,\beta)
\label{Family_General_Heisenberg:eq}
\end{equation}
for all $a,b > 0$, $\alpha > \frac{d}{d+a}$, $\beta > \frac{d}{d+b}$, $\beta \le
\frac{\alpha}{2  \alpha-1}$  and  with  the   bounds  $\Z$  and  $\B$  given  in
eqs.~\eqref{BoundZPV:eq} and~\eqref{BoundM:eq}.


\subsection{The maximal bound is on the conjugation curve $\beta = \alpha^*$}
\label{MaxBoundConj:app}

We will now show that the pair $(\alpha,\beta)$ that maximizes $\Z(\alpha,\beta)
\M(a,\alpha)  \M(b,\beta)$ is  on the  conjugation  curve, namely  for $\beta  =
\alpha^* = \alpha  / (2 \alpha - 1)$,  for any values of $a$ and  $b$ (under the
existence condition for $\M$).


\subsubsection{Function $\M(l,\lambda)$ is increasing with $\lambda$}
\label{MIncreases:app}

Let us first consider the derivative of $\M(l,\lambda)$ versus $\lambda$.

For $\lambda > 1$, i.e.\ $\mu = \frac{\lambda}{\lambda-1} > 1$,
\begin{eqnarray*}
\frac{\partial}{\partial \mu} \log \M & = & \frac{\partial}{\partial \mu}
\left(- \frac{2}{d} \log \Gamma(\mu) + \frac{2}{d} \log \Gamma \left( \mu +
\frac{d}{l} \right) - \frac{2}{l} \log(d + l \mu) + \frac{2 (\mu-1)}{d} \log
\left( \frac{l \mu}{d + l \mu} \right) \right)\\
& = & \frac{2}{d} \left( - \psi(\mu) + \psi \left( \mu + \frac{d}{l} \right) +
\frac{l}{d + l \mu} - \frac{1}{\mu} + \log \left( \frac{l \mu}{d + l \mu}
\right) \right)
\end{eqnarray*}
where  $\psi(x) =  \frac{d}{d x}  \log\Gamma(x)$ is  the digamma  function.

Similarly, for $\lambda \in \left( 1  - \frac{l}{l+d} ; 1 \right)$, i.e.\ $\mu <
- d/l$,
\begin{eqnarray*}
\frac{\partial}{\partial \mu} \log \M & = & \frac{\partial}{\partial \mu} \left(
- \frac{2}{d} \log \Gamma \left( 1 - \mu - \frac{d}{l} \right) + \frac{2}{d}
\log\Gamma(1-\mu) \right.\\
& & \left. - \frac{2}{l} \log( - d - l \mu) + \frac{2
(\mu-1)}{d} \log \left( \frac{l \mu}{d + l \mu} \right) \right)\\
& = & \frac{2}{d} \left( - \psi(1-\mu) + \psi \left( 1 - \mu - \frac{d}{l}
\right) + \frac{l}{d + l \mu} - \frac{1}{\mu} + \log \left( \frac{l \mu}{d + l
\mu} \right) \right)\\
& = & \frac{2}{d} \left( \psi \left(- \frac{d + l \mu}{l} \right) - \psi(-\mu) +
\log \left( \frac{l \mu}{d + l \mu} \right) \right)
\end{eqnarray*}
the last simplification coming from~\cite[Eq. 8.365-1]{GraRyz07}.

To summarize, noting that $\partial \mu / \partial \lambda = -1/(\lambda-1)^2$,
\begin{equation}\left\{\begin{array}{lllll}
\frac{\partial}{\partial \lambda} \log \M(l,\lambda) & = & \displaystyle
\frac{2}{d (\lambda-1)^2} \left( \psi(\mu) + \frac{1}{\mu} - \log \mu \right.\\
& & \hspace{2cm} \displaystyle \left.- \psi
\left(\mu + \frac{d}{l} \right) - \frac{1}{\mu + \frac{d}{l}} + \log \left( \mu
+ \frac{d}{l} \right) \right) & \mbox{ if } & \lambda > 1\vspace{5mm}\\
\frac{\partial}{\partial \lambda} \log \M(l,\lambda) & = & \displaystyle
\frac{2}{d (\lambda-1)^2} \left( \psi(-\mu) -\log(-\mu) \right.\\
& & \hspace{2cm} \displaystyle \left. - \psi \left(- \mu -
\frac{d}{l} \right) + \log \left( -\mu - \frac{d}{l} \right) \right) & \mbox{ if
} & 1 - \frac{l}{l+d} < \lambda < 1
\end{array}\right.\label{DerivativeLogM:eq}\end{equation}
Taking the  limit $\lambda \to  1^+$ from the  first line, or $\lambda  \to 1^-$
from the  second line, and  $\M$ being continuous  in $\lambda = 1$,  we achieve
$\left.\frac{\partial}{\partial \lambda} \log \M(l,\lambda)\right|_{\lambda=1} =
\frac{1}{l}$.

Let us consider now  the terms in the parentheses in the  right-hand side of the
first line  in Eq. \eqref{DerivativeLogM:eq}. They  can be written  as $g(\mu) -
g(\mu+d/l)$ with,
\begin{equation}
g(\mu) = \psi(\mu)+\frac{1}{\mu}-\log \mu.
\label{FunctionG:eq}
\end{equation}
Then, from~\cite[6.4.1]{AbrSte70}
\begin{eqnarray*}
g'(\mu) & = & \psi'(\mu) - \frac{1}{\mu} - \frac{1}{\mu^2}\\
& = & \int_0^{+\infty} \frac{t}{1-e^{-t}} e^{- \mu t} \d t - \int_0^{+\infty} e^{-
\mu t} \d t - \int_0^{+\infty} t \, e^{- \mu t} \d t\\
& = & \int_0^{+\infty} \frac{-1+e^{-t}+t e^{-t}}{1-e^{-t}} e^{- \mu t} \d t.
\end{eqnarray*}
Now, it  is easy  to show that  $-1+e^{-t}+t e^{-t}  \le 0$ for  $t \ge  0$ that
permits to  conclude that  $g' \le  0$ and thus  that $g$  is decreasing.   As a
conclusion,  $g(\mu)  - g(\mu+d/l)  \ge  0$  and thus  $\frac{\partial}{\partial
  \lambda} \log \M \ge 0$: $\M$ is increasing in $(1 \: ; \: +\infty)$.

Similarly, the terms in parentheses in  the right-hand side of second line ($\mu
< -d/l < 0$ here), writes $h(\mu) - h(\mu+d/l)$ with,
\begin{eqnarray*}
h(\mu) = \psi(-\mu)-\log(-\mu)
\end{eqnarray*}
and give from~\cite[6.4.1]{AbrSte70}
\begin{eqnarray*}
h'(\mu) & = & - \psi'(-\mu) - \frac{1}{\mu}\\
& = & \int_0^{+\infty} \frac{-t}{1-e^{-t}} e^{\mu t} \d t + \int_0^{+\infty}
e^{\mu t} \d t\\
& = & \int_0^{+\infty} \frac{1-t-e^{-t}}{1-e^{-t}} e^{\mu t} \d t.
\end{eqnarray*}
Then, it is easy  to show that $1-t-e^{-t} \le 0$ for $t  \ge 0$ that permits to
conclude that  $h' \le  0$ and thus  that $h$  is decreasing.  As  a conclusion,
$h(\mu)  -   h(\mu+d/l)  \ge   0$  and  thus   also  for  $\lambda   \in  \left(
  1-\frac{l}{l+d} \: ; \:  1 \right)$ we have $\frac{\partial}{\partial \lambda}
\log \M \ge 0$: $\M$ is increasing.


\subsubsection{$\B(\lambda)$  increases with  $ \lambda  \in [1/2  \: ;  1]$ and
  decreases with $ \lambda > 1$}
\label{BBehavior:app}

From~\eqref{Bound_Conj:eq}  and  $\lambda^*  =  \lambda /  (2  \lambda-1)$,  the
derivative of $\B(\lambda)$ writes
\begin{eqnarray*}
\frac{\partial}{\partial \lambda} \log \B(\lambda) & = &
\frac{\partial}{\partial \lambda} \left( \frac{\log \lambda}{\lambda-1} +
\frac{\log \lambda^*}{\lambda^*-1} \right)\\
& = & \frac{\partial}{\partial \lambda} \left( \frac{\log \lambda}{\lambda-1}
\right) + \frac{\partial}{\partial \lambda^*} \left( \frac{\log
\lambda^*}{\lambda^*-1} \right) \frac{\partial \lambda^*}{\partial \lambda}\\
& = & \left( \frac{1}{\lambda (\lambda-1)} - \frac{\log \lambda}{(\lambda-1)^2}
\right) - \left( \frac{1}{\lambda^* (\lambda^*-1)} - \frac{\log
\lambda^*}{(\lambda^*-1)^2} \right) \frac{1}{(2 \lambda-1)^2}
\end{eqnarray*}
that is
\begin{equation}
\frac{\partial}{\partial \lambda} \log \B(\lambda) = \frac{1}{
(\lambda-1)^2} \left( 2 - \frac{2}{\lambda} - \log (2 \lambda-1) \right).
\end{equation}
A short study of  the right hand side, show that this  quantity is positive if $
\lambda \in [1/2 \:  ; 1]$ and negative if $\lambda \ge  1$: $\B$ increases with
$\lambda$ in $[1/2 \: ; 1]$ and then decreases for larger values of $\lambda$.


\subsubsection{Domain where the maximal bound has to be searched}
\label{DomainMaxBound:app}

\begin{figure}[htbp]
\centerline{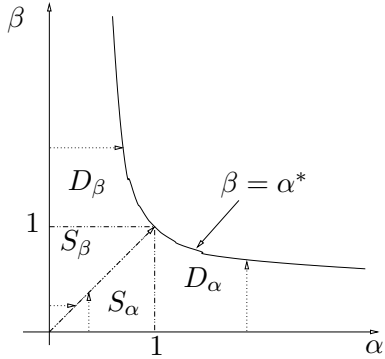}
\caption{Sets  $D_\alpha$,  $D_\beta$, $S_\alpha$  and  $S_\beta$  in the  plane
  $(\alpha,\beta)$.   The  solid  curve   represents  the  pairs  of  conjugated
  parameters,  i.e.   $\beta  =  \alpha^*$.   The dotted  arrows  indicate  that
  $\Z(\alpha,\beta)   \M(a,\alpha)   \M(b,\beta)$   increases  when   the   pair
  $(\alpha,\beta)$  moves along  their directions,  in the  sets where  they are
  plotted.}
\label{Sets:fig}
\end{figure}

Remind that starting from~\eqref{Family_General_Heisenberg:eq}, namely $ \langle
r^a\rangle^{\frac{2}{a}}    \,    \langle    p^b    \rangle^{\frac{2}{b}}    \ge
\Z(\alpha,\beta)  \M(a,\alpha) \M(b,\beta)$,  the  best bound  is  then so  that
$\Z(\alpha,\beta)  \M(a,\alpha)  \M(b,\beta)$  is  maximized as  a  function  of
$\alpha$  and  $\beta$.   Let  us  now  consider  the   following  sets  in  the
$(\alpha,\beta)$ plane:
\begin{eqnarray*}\left\{\begin{array}{lll}
D_\alpha & = & \left\{ \left. (\alpha,\beta) \in \Rset_+^2 \right| \alpha \ge 1,
\beta \le \alpha^* \right\}\vspace{2mm}\\
D_\beta & = & \left\{ \left. (\alpha,\beta) \in \Rset_+^2 \right| \beta \ge 1,
\alpha \le \beta^* \right\}\vspace{2mm}\\
S_\alpha & = & \left\{ \left. (\alpha,\beta) \in [0 \: ; \: 1]^2 \right| \beta
\le \alpha \right\}\vspace{2mm}\\
S_\beta & = & \left\{ \left. (\alpha,\beta) \in [0 \: ; \: 1]^2 \right| \alpha
\le \beta \right\}\vspace{2mm}\\
S_1 & = & S_\alpha \cup S_\beta
\end{array}\right.\end{eqnarray*}
where  $\alpha^*  = \frac{\alpha}{2  \alpha-1}$  and  $\beta^* =  \frac{\beta}{2
  \beta-1}$. These sets are represented in Fig.~\ref{Sets:fig}.

In order to study the best bound, we consider each subset:
\begin{itemize}
\item   We  first   consider   domain  $D_\alpha$   and   fix  $\alpha$.    From
  Eq.  \eqref{BoundZPV:eq},  the bound  is  then\\ $\Z(\alpha,\beta)  \M(a,\alpha)
  \M(b,\beta) = \B(\alpha) \M(a,\alpha) \M(b,\beta)$ and from the previous study
  of $\M$ we can know that it increases with $\beta$. Thus, the bound is maximum
  precisely {\em on} the conjugation curve $\beta = \alpha^*$.
\item By symmetry, in domain $D_\beta$  and fixing $\beta$, one shows again that
  the bound is maximal on the conjugation curve $\alpha = \beta^*$.
\item  In the domain $S_1$ we discuss the following cases:
  \begin{itemize}
  \item The maximum bound must be achieved on the line segment $\alpha = \beta$.
    Indeed,  in  $S_\alpha$, the  bound  is  given  by $\B(\alpha)  \M(a,\alpha)
    \M(b,\beta)$  if   $\alpha  \ge  1/2$   and  $\M(a,\alpha)  \M(b,\beta)/e^2$
    otherwise. Again, fixing $\alpha$, the  bound is increasing with $\beta$ and
    thus is  maximum for $\beta =  \alpha$. This remains valid,  by symmetry, in
    $S_\beta$, and thus in all $S_1$.
  \item For $\alpha \le 1/2$, on the  line segment $\alpha = \beta$ the bound is
    $\M(a,\alpha)  \M(b,\alpha)/e^2$ and  thus  increases with  $\alpha$: it  is
    maximum for $\alpha = 1/2$.
  \item  For $\alpha  \in (1/2  \:  ; \:  1]$, the  bound expresses  $\B(\alpha)
    \M(a,\alpha) \M(b,\alpha)$ and  tends to $\M(a,\alpha) \M(b,\alpha)/e^2$ for
    $\alpha \to  1/2$. $\B$ being an increasing  function in $[1/2 \:  ; 1]$ and
    since $\M$ is increasing, the bound is then maximum for $\alpha=1$.
  \end{itemize}
  As a conclusion, on $S_1$ the maximum bound is achieved when $\alpha = \beta =
  1$ that is again {\em on} the conjugation curve.
\end{itemize}


\subsection{Maximal bound and properties}
\label{MaximalBound:app}

The best  bound of the  generalized Heisenberg relation  one can achieve  by our
approach is then
\begin{equation}
\C(a,b) = \max_{\alpha \in D(a,b)} \B(\alpha) \M(a,\alpha) \M(b,\alpha^*)
\end{equation}
where the  domain of search  $D$ is  ruled by the  restriction on the  domain of
existence of $\M$. We will come back later on this domain.


\subsubsection{Symmetries}
\label{Symmetries:app}

Let us denote by $\alpha_{\mathrm{opt}}(a,b)$ the index that leads to $\C(a,b)$,
i.e.
\begin{equation}
\alpha_{\mathrm{opt}}(a,b) = \arg\max_\alpha \B(\alpha) \M(a,\alpha) \M(b,\alpha^*)
\label{AlphaOpt:eq}
\end{equation}

Noticing  that  $\B(\alpha)  =   \B(\alpha^*)$,  one  immediately  observes
from~\eqref{AlphaOpt:eq} that
\begin{equation}\left\{\begin{array}{lll}
\C(b,a) & = & \C(a,b)\vspace{2mm}\\
\alpha_{\mathrm{opt}}(b,a) & = & (\alpha_{\mathrm{opt}}(a,b))^*.
\end{array}\right.\end{equation}

Thus, without loss of generality, one can restrict the study to the case with $a
\ge b$.


\subsubsection{Reduced domain of search}
\label{ReducedDomain:app}

Consider the situation where $a \ge b$.

If $\alpha > 1$,  then $\alpha^* < 1$.  We will show  that the bound $\B(\alpha)
\M(a,\alpha)  \M(b,\alpha^*)$ decreases  with  $\alpha$; thus  the maximum  must
satisfy $\alpha \le 1$.

We have  already seen  that $\B(\alpha)$ decreases  when $\alpha >  1$. Consider
then  the  part  $\M(a,\alpha)  \M(b,\alpha^*)$. Remembering  that  $\alpha^*  =
\frac{\alpha}{2 \alpha-1}$, one has $\frac{\partial \alpha^*}{\partial \alpha} =
-  \frac{1}{(2 \alpha  -1)^2}$.  Moreover, one  has $\frac{1}{(\alpha^*-1)^2}  =
\frac{(2       \alpha-1)^2}{(\alpha-1)^2}$       and      $\mu(\alpha^*)       =
\frac{\alpha^*}{\alpha^*-1}  = -  \frac{\alpha}{\alpha-1}  = -\mu(\alpha)$  from
Eq. \eqref{mu:eq}. Then, from~\eqref{DerivativeLogM:eq},
\begin{eqnarray*}
\frac{\partial}{\partial \alpha} \log \left(\M(a,\alpha) \M(b,\alpha^*)
\right) & = & \frac{\partial}{\partial \alpha} \log \M(a,\alpha) +
\frac{\partial \alpha^*}{\partial \alpha} \frac{\partial}{\partial \alpha^*}
\log \M(b,\alpha^*)\vspace{2mm}\\
& \hspace{-6cm} = & \hspace{-3cm} \frac{2}{d (\alpha-1)^2} \left(\frac{1}{\mu} -
\frac{1}{\mu+d_a} - \psi(\mu+d_a) + \psi(\mu-d_b) + \log(\mu+d_a) -
\log(\mu-d_b) \right)
\end{eqnarray*}
where $\mu$ stands for  $\mu(\alpha)$, $d_a = d/a$ and $d_b =  d/b$. The goal is
then to show the negativity of
\begin{equation}
k(\mu,d_a,d_b) = \frac{1}{\mu} - \frac{1}{\mu+d_a} - \psi(\mu+d_a) +
\log(\mu+d_a) + \psi(\mu-d_b)  - \log(\mu-d_b)
\end{equation}
keeping in mind  that $d_a \le d_b$. To  this end, we can view  this function in
terms of $d_a$ for instance, and  thus the sense of variation of $k(\mu,d_a,d_b)
= -g(\mu+d_a) + g(\mu-d_b) + 1/\mu -  1/(\mu-d_b)$ is the same than the sense of
variation   of  $-   g(\mu+d_a)  =   -  \psi(\mu+d_a)   -   \frac{1}{\mu+d_a}  +
\log(\mu+d_a)$ introduced Eq.~\eqref{FunctionG:eq}, versus $d_a$.  We have shown
that function  $g$ is decreasing and  thus $k$ is increasing  with $d_a$. Since,
$d_a \le d_b$, to show that $k$  is negative, it is then sufficient to show that
$k(\mu,d_b,d_b)$ is negative.  From~\cite[6.4.1]{AbrSte70}
\begin{eqnarray*}
\frac{\partial k(\mu,d_b,d_b)}{\partial d_b} & = & \frac{1}{\mu+d_b} +
\frac{1}{(\mu+d_b)^2} - \psi'(\mu+g_g) + \frac{1}{\mu-d_b} -
\psi'(\mu-d_b)\\[2.5mm]
& = & \int_0^{+\infty} \left[ \left( 1 + t - \frac{t}{1-e^{-t}} \right) e^{- d_b
t} + \left( 1 - \frac{t}{1-e^{-t}} \right) e^{+ d_b t} \right] e^{- \mu t} \d
t\\[2.5mm]
& = & \int_0^{+\infty} \left[ \left( 1 - e^{-t} - t e^{-t} \right) e^{- d_b t} +
\left( 1 - t - e^{-t} \right) e^{+ d_b t} \right] \frac{e^{- \mu t}}{1-e^{-t}}
\d t.
\end{eqnarray*}
Now, it  is quite easy to  show that the  term in square brackets  is decreasing
with $d_b$ since the derivative in $d_b$ is negative (the factors of $e^{\pm d_b
  t}$ are negative). For  $d_b = 0$, it is no more hard  to show that the square
bracket is  negative, that  permits to  conclude that for  any $d_b$  the square
bracket term is negative: $k(\mu,d_b,d_b)$ decreases with $d_b$.

At last $k(\mu,0,0) = 0$ and  thus $k(\mu,d_b,d_b) \le 0$, implying that for any
$d_a \le d_b$ one has $k(\mu,d_a,d_b) \le 0$.

As   claimed,    $\frac{\partial}{\partial   \alpha}   \log   \left(\M(a,\alpha)
  \M(b,\alpha^*) \right) \le 0$ for $\alpha  > 1$. Together with the decrease of
$\B$ when $\alpha > 1$ we  conclude that the maximum of $\B(\alpha) \M(a,\alpha)
\M(b,\alpha^*)$ is attained for $\alpha < 1$.

To finish to determine  the domain of search for the maximal  bound, when $a \ge
b$, implies that $\alpha  \le 1$ as just found, $\alpha >  1/2$ since it must by
on  the   conjugation  curve,  and   from~\eqref{DerivativeLogM:eq},  $\alpha  >
\frac{d}{d+a}$.

\

In summary, for $a \ge b$,
\begin{equation}
\C(a,b) = \max_{\alpha \in \left( \max \left( \frac12 , \frac{d}{d+a} \right) \: ; \: 1 \right]}
\B(\alpha) \M(a,\alpha) \M(b,\alpha^*)
\end{equation}
where   $\B$   and   $\M$   are  respectively   given   by~\eqref{Bound_Conj:eq}
and~\eqref{BoundM:eq}.


\bibliography{HeisenbergMomentosUR}

\end{document}

%% file: FIG_Sets.pdf_t
\begin{picture}(0,0)%
\includegraphics{FIG_Sets.pdf}%
\end{picture}%
\setlength{\unitlength}{1450sp}%
\begingroup\makeatletter\ifx\SetFigFont\undefined%
\gdef\SetFigFont#1#2#3#4#5{%
  \reset@font\fontsize{#1}{#2pt}%
  \fontfamily{#3}\fontseries{#4}\fontshape{#5}%
  \selectfont}%
\fi\endgroup%
\begin{picture}(6372,6138)(1516,-5962)
\put(2431,-4066){\makebox(0,0)[lb]{\smash{{\SetFigFont{6}{7.2}{\familydefault}{\mddefault}{\updefault}{\color[rgb]{0,0,0}{\small $S_\beta$}}%
}}}}
\put(7651,-5821){\makebox(0,0)[lb]{\smash{{\SetFigFont{6}{7.2}{\familydefault}{\mddefault}{\updefault}{\color[rgb]{0,0,0}{\small $\alpha$}}%
}}}}
\put(1531,-241){\makebox(0,0)[lb]{\smash{{\SetFigFont{6}{7.2}{\familydefault}{\mddefault}{\updefault}{\color[rgb]{0,0,0}{\small $\beta$}}%
}}}}
\put(1846,-3751){\makebox(0,0)[lb]{\smash{{\SetFigFont{6}{7.2}{\familydefault}{\mddefault}{\updefault}{\color[rgb]{0,0,0}{\small $1$}}%
}}}}
\put(5176,-3076){\makebox(0,0)[lb]{\smash{{\SetFigFont{6}{7.2}{\familydefault}{\mddefault}{\updefault}{\color[rgb]{0,0,0}{\small $\beta = \alpha^*$}}%
}}}}
\put(3961,-5866){\makebox(0,0)[lb]{\smash{{\SetFigFont{6}{7.2}{\familydefault}{\mddefault}{\updefault}{\color[rgb]{0,0,0}{\small $1$}}%
}}}}
\put(2566,-3031){\makebox(0,0)[lb]{\smash{{\SetFigFont{6}{7.2}{\familydefault}{\mddefault}{\updefault}{\color[rgb]{0,0,0}{\small $D_\beta$}}%
}}}}
\put(3241,-5101){\makebox(0,0)[lb]{\smash{{\SetFigFont{6}{7.2}{\familydefault}{\mddefault}{\updefault}{\color[rgb]{0,0,0}{\small $S_\alpha$}}%
}}}}
\put(4546,-4696){\makebox(0,0)[lb]{\smash{{\SetFigFont{6}{7.2}{\familydefault}{\mddefault}{\updefault}{\color[rgb]{0,0,0}{\small $D_\alpha$}}%
}}}}
\end{picture}%